\documentclass[lettersize,journal]{IEEEtran}
\usepackage{amsmath,amsfonts}
\usepackage{algorithmic}
\usepackage{algorithm}
\usepackage{array}
\usepackage[caption=false,font=normalsize,labelfont=sf,textfont=sf]{subfig}
\usepackage{textcomp}
\usepackage{stfloats}
\usepackage{url}
\usepackage{verbatim}
\usepackage{graphicx}
\usepackage{cite}
\usepackage{bm}
\usepackage {booktabs}
\usepackage{url}
\usepackage[implicit=false]{hyperref}
\begin{document}
\title{
 Computer Vision-Aided Reconfigurable Intelligent Surface-Based Beam Tracking: Prototyping and Experimental Results }


\author{Ming~Ouyang, Yucong~Wang, Feifei~Gao,~\IEEEmembership{Fellow,~IEEE,} Shun~Zhang,~\IEEEmembership{Senior Member,~IEEE,} Puchu~Li, and~Jian~Ren
	\thanks{M.~Ouyang, Y.~Wang and F.~Gao are with the Department of Automation, Tsinghua University, Beijing 100084, China, and also with Beijing National Research Center for Information Science and Technology (BNRist), Beijing 100084, China (email: \protect\href{mailto:oym21@mails.tsinghua.edu.cn}{oym21@mails.tsinghua.edu.cn};
		\protect\href{mailto:wangyuco21@mails.tsinghua.edu.cn}{wangyuco21@mails.tsinghua.edu.cn}; \protect\href{mailto:feifeigao@ieee.org}{feifeigao@ieee.org}).

	 S.~Zhang is with the State Key Laboratory of Integrated Services Networks, Xidian University, Xi’an 710071, China (e-mail:\protect\href{mailto:zhangshunsdu@xidian.edu.cn}{zhangshunsdu@xidian.edu.cn}).

	 P.~Li and~J.~Ren are with the National Key Laboratory of Antennas and Microwave Technology, Xidian University, Xi’an 710071, China (e-mail:\protect\href{mailto:puchuli@163.com}{puchuli@163.com};
	 \protect\href{mailto:renjian@xidian.edu.cn}{renjian@xidian.edu.cn}).
	}
   
	}

\markboth{}%
{111}

\maketitle

\begin{abstract}
 In this paper, we propose a novel computer vision-based approach to aid Reconfigurable Intelligent Surface (RIS) for dynamic beam tracking and then implement the corresponding prototype verification system. A camera is attached at the RIS to obtain the visual information about the surrounding environment, with which RIS identifies the desired reflected beam direction and then adjusts the reflection coefficients according to the pre-designed codebook. Compared to the conventional approaches that utilize channel estimation or beam sweeping to obtain the reflection coefficients, the proposed one not only saves beam training overhead but also eliminates the requirement for extra feedback links.
 We build a 20-by-20 RIS running at 5.4 GHz and develop a high-speed control board to ensure the real-time refresh of the reflection coefficients. Meanwhile we implement an independent peer-to-peer communication system to simulate the communication between the base station and the user equipment.  The vision-aided RIS prototype system is tested in two mobile scenarios: RIS works in near-field conditions as a passive array antenna of the base station; RIS works in far-field conditions to assist the communication between the base station and the user equipment. The experimental results show that RIS can quickly adjust the reflection coefficients for dynamic beam tracking with the help of visual information. 
\end{abstract}


\section{Introduction}

\IEEEPARstart{R}{econfigurable} intelligent surface (RIS) assisted communications has emerged as one of the main technologies for the next-generation mobile communication systems and can deliver significant improvements at a cheap cost \cite{9253607,9679804,9387701,9086766,9424177}. A RIS is a uniform array of elements that can modulate the incident wave's amplitude and phase. The primary principle of RIS is to leverage the array elements' modulation abilities to generate certain reflected beams, thus enhancing the signal power in the specific directions. The key challenge in applying RIS to assist communications is how to design the appropriate reflection coefficients for dynamic beam tracking. Currently, there are three main ways to compute the reflection coefficients: channel state information (CSI)-based schemes, beam-sweeping-based schemes, and end-to-end deep learning-based schemes.

The CSI-based schemes can be further divided into perfect CSI-based \cite{trabeamforming1,trabeamforming8,trabeamforming9,trabeamforming2,trabeamforming3,9594786} and statistical CSI-based schemes \cite{trabeamforming7,trabeamforming6,trabeamforming11}. In \cite{trabeamforming1}, the authors selected the optimal reflection coefficients from a pre-designed codebook after estimating the overall equivalent channel such that the spectral efficiency can be maximized. In \cite{trabeamforming8}, the authors designed a low overhead majorization-minimization-based method to optimize the reflection coefficients. In \cite{trabeamforming9}, the authors divided the channel into multiple subchannels, each corresponding to a RIS array element, and calculated the reflection coefficients based on the CSI of each subchannel. In \cite{trabeamforming2,trabeamforming3,9594786}, the authors introduced compressed sensing into the channel estimation process and calculated the reflection coefficients based on the recovered sparse CSI. Typically, the statistical CSI is easier to obtain than the perfect CSI. In \cite{trabeamforming7,trabeamforming6,trabeamforming11}, the authors optimized the reflection coefficients based on the statistical CSI and analyzed the performance of the optimization algorithms. Although it is very effective to design reflection coefficients based on the CSI, the huge number of elements on RIS leads to a large training overhead.

Compared with the CSI-based schemes, the beam-sweeping-based schemes does not require complex channel estimation and therefore  are highly advantageous for RIS with a large number of array elements \cite{9473674,trichopoulos2021design,9406940}. In \cite{9473674}, the authors utilized a manifold-based algorithm to calculate the optimal reflection coefficients for specific directions and then took beam-sweeping scheme to obtain the correct direction. In \cite{trichopoulos2021design,9406940}, the authors validated the beam-sweeping scheme in the real-world environment. In order to speed up the sweep, the authors of \cite{9551980} introduced a greedy algorithm to search for the optimal reflection coefficients. However, these beam-sweeping schemes not only consumes a large amount of sweep time, but also requires extra feedback links, which increases the communication delay.

 In end-to-end deep learning-based schemes, \cite{9317827} and \cite{9580332} leveraged neural networks to output the optimal reflection coefficients directly. However, the neural network in \cite{9317827} requires the user's location information as input, and the neural network in \cite{9580332} requires the power distribution near the user as input. The acquisition of the location information or the power distribution likewise results in significant additional training overhead.

 Recently, out-of-band information is introduced into the communication systems to reduce the beam training overhead and eliminate the feedback links, e.g., \cite{frehop,radarcom,vision1,vision2,vision3,vision4}. In \cite{frehop}, sub-6 GHz channel covariance was applied to assist in estimating the mmWave channel covariance.  In \cite{radarcom}, MIMO radar was utilized to aid mmWave base station in estimating the time-varying channel.
 In \cite{vision1,vision2,vision3,vision4}, visual information was leveraged to assist base station in realizing beam tracking, blockage prediction or proactive handoff. For RIS, applying visual information to design reflection coefficients may be a much more suitable option, due to the following three reasons: (1) RIS is mainly used to assist communication through reflected line of sight (LOS) path, especially for mmWave band and Terahertz band, whereas the camera is effective precisely when used to quickly locate users within LOS; (2) One of the advantages of RIS is low cost, therefore it is appropriate to replace the complex feedback links with a cheap camera; (3) The acquisition of visual information does not occupy other frequency bands, which saves spectrum resources.

In this paper, we propose a novel computer vision-based approach to aid RIS for dynamic beam tracking and implement the corresponding prototype verification system. Compared to the conventional schemes that utilize channel estimation or beam sweeping, the proposed one not only saves beam training overhead but also eliminates the requirement for extra feedback links. The main contributions of this paper can be summarized as follows:
\begin{itemize}
	\item{For the first time, visual information is employed to assist RIS in realizing beam tracking. The proposed vision-based scheme utilizes advanced computer vision technology to acquire the direction of the user relative to the RIS, and then selects the near-optimal reflection coefficients in the pre-designed codebook.}

	\item{A high-speed control board is developed to realize the real-time refresh of reflection coefficients. We cascaded two low-cost, high-I/O-density FPGA chips (Intel Cyclone IV EP4CE15F23C8N) in the control board, which can run at up to 200MHz system clock frequency and control each PIN diode individually.}
	\item{The proposed prototype system is tested in two classical scenarios: RIS works in near-field conditions as a passive array antenna of the base station; RIS works in far-field conditions to assist the communication between the base station and the user equipment. The experimental results show that RIS can quickly adjust the reflection coefficients for dynamic beam tracking with the help of visual information.}
\end{itemize}

The remainder of the paper is composed of the following parts: Section \uppercase\expandafter{\romannumeral2} presents the system model and the design principle of the reflection coefficients. Section \uppercase\expandafter{\romannumeral3} provides the specific implementation details of vision-based beam tracking scheme. Section \uppercase\expandafter{\romannumeral4} describes the vision-aided RIS prototype verification system, including the design of the RIS codebook, the control board, and the architecture of the peer-to-peer communication system. In Section \uppercase\expandafter{\romannumeral5}, we test the vision-aided RIS in two scenarios and analyze the benefits of visual information. Finally, in Section \uppercase\expandafter{\romannumeral6}, we make the conclusions.
\section{System Model} 
\begin{figure}[!t]
	\centering
	\includegraphics[width=2.5in]{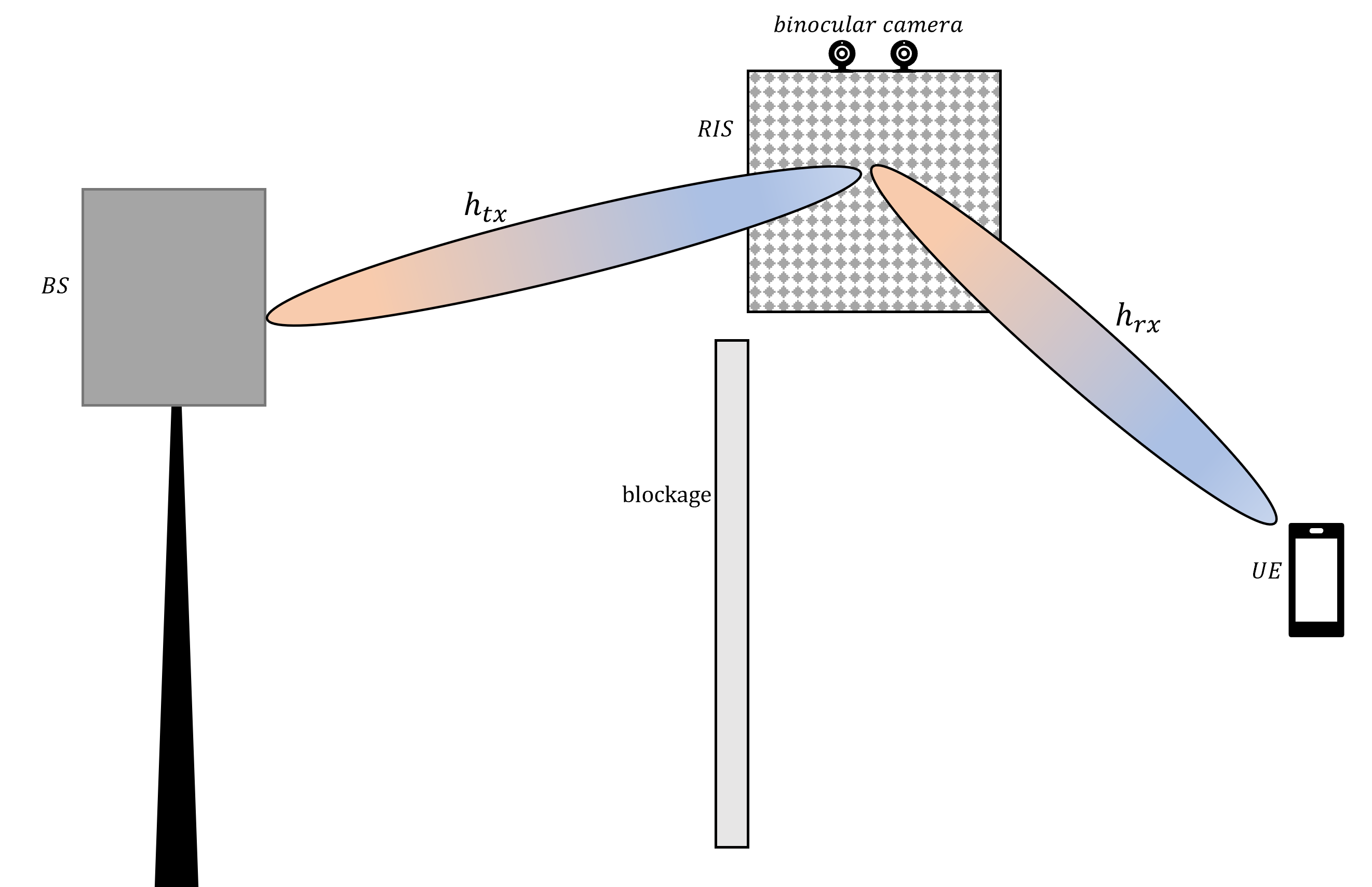}
	\caption{Architecture diagram of the vision-aided RIS prototype system, in which the light of sight (LOS) path between the BS and the UE is blocked.}
	\label{situfig_1}
\end{figure}

  We assume that the base station (BS) and RIS serve only one user equipment (UE) in the communication system, as shown in Fig.~\ref {situfig_1}. 
  Moreover, the BS and UE are equipped with single antenna, while the RIS contains $ M \times N $ units. The orthogonal frequency division multiplexed (OFDM) is adopted and the baseband contains $ K $ subcarriers. Let $\bm{h}_{rx}[k],\bm{h}_{tx}[k]\in\mathbb{C}^{MN\times1}$ represent the channel between the UE and RIS, and that between the BS and RIS at the $k$-th subcarrier, respectively. Denote $\bm{H}_{RIS}\in\mathbb{C}^{MN \times MN}$ as the manipulation matrix at RIS. Note that $ \bm{H}_{RIS} $ is a diagonal matrix, whose diagonal elements can form the vector $ \bm{h}_{ris}=[A_{11}e^{j\alpha_{11}},\cdots,A_{1N}e^{j\alpha_{1N}},\cdots,A_{MN}e^{j\alpha_{MN}}]^T $, where $ A_{mn},\alpha_{mn} $ represent the modulated phase and modulated amplitude of the $mn$-th unit in RIS. We assume the line of sight (LOS) path between BS and UE is blocked, and then the UE's received signal $ r_k $ at the $k$-th subcarrier can be written as
\begin{equation}
	\label{model_ris1}
	r_k = \bm{h}^T_{rx}[k]\bm{H}_{RIS}\bm{h}_{tx}[k]s_k+n_k ,
\end{equation}
where  $ s_k,n_k\in\mathbb{C} $ denote the BS's transmitting signal and the noise. The channel capacity $C$ can be computed as
\begin{equation}
	\label{model_ris2}
	C =\sum_{i=1}^{K}\log_{2} {\left(1+\dfrac{|\bm{h}^T_{rx}[k]\bm{H}_{RIS}\bm{h}_{tx}[k]s_k|^2P_k}{\sigma^2}\right)} ,
\end{equation}
where $ P_k,\sigma^2 $ represent the average power of the signal and the variance of the noise, respectively.

A optimal manipulation matrix $ \bm{H}_{RIS} $ should be designed to to maximize the channel capacity $ C $. According to \eqref{model_ris2}, we can obtain the approximate optimal beamforming vector $ \bm{h}_{ris}^{optm} $ as
\begin{equation}
	\label{model_ris3}
	\bm{h}_{ris}^{optm} =\mathop{\arg\max}\limits_{A_{mn},\alpha_{mn}}\ \prod_{i=1}^{K}|\bm{h}^T_{rx}[k]\text{diag}(\bm{h}_{ris})\bm{h}_{tx}[k]s_k|^2P_k .
\end{equation}
According to \eqref{model_ris3}, the prerequisite for calculating $ \bm{h}_{ris}^{optm} $ is to obtain the CSI $\bm{h}_{rx}[k]$ and $\bm{h}_{tx}[k]$. However, the large number of array units on the RIS poses a great challenge for channel estimation.

\begin{figure}[!t]
	\centering
	\includegraphics[width=2.5in]{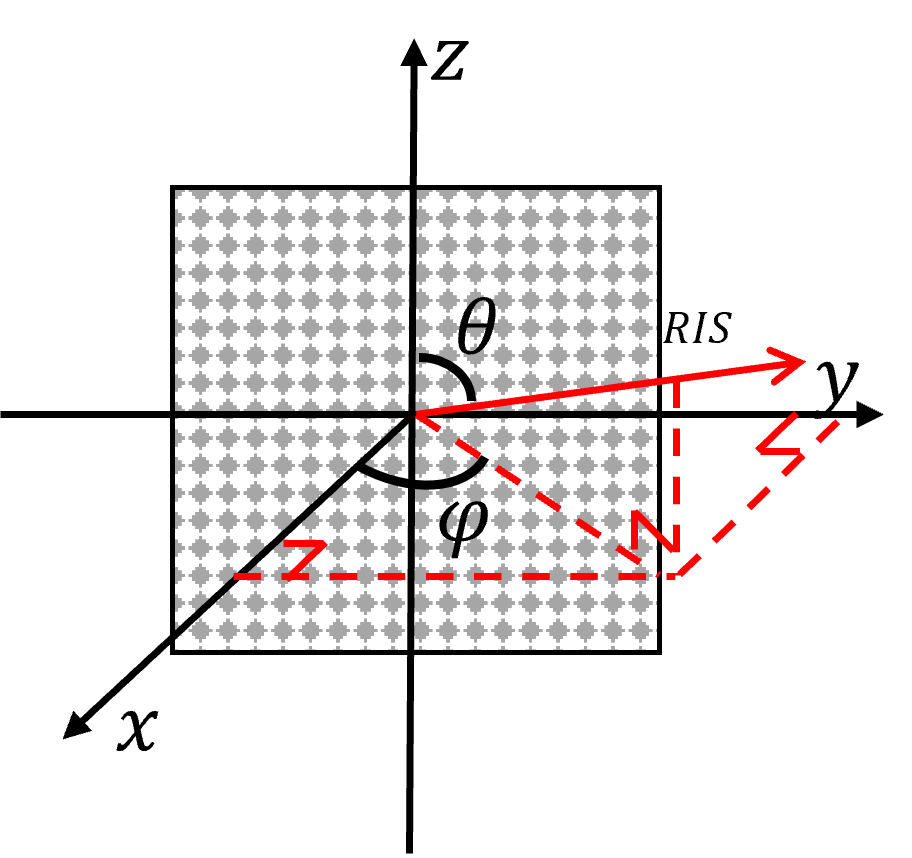}
	\caption{Coordinate system established with the RIS center as the origin, where $ \theta$ and $\varphi $ are the pitch and azimuth angles relative to RIS, respectively.}
	\label{situfig_2}
\end{figure}

In order to deal with the challenge of channel estimation, we transform \eqref{model_ris3} into maximizing the received signal power at the UE. We estabilish a coordinate system with the center point of the RIS board as the coordinate origin and use it as the phase reference point, as shown in Fig.~\ref {situfig_2}.
According to \cite{2016A}, the scattering field pattern function of RIS can be expressed as 
\begin{gather}
	\label{model_ris4}
    E(\theta,\varphi)=\sum_{m=1}^{M}\sum_{n=1}^{N}f_{mn}(\theta,\varphi)A_{mn}B_{mn}exp\left(j(\alpha_{mn}+\beta_{mn})\right) \notag \\
	exp\left(j\left(\dfrac{2\pi d}{\lambda}\left(\left(\dfrac{M+1}{2}-m\right)\cos\theta\right)\right)\right)
	\notag \\
	exp\left(j\left(\dfrac{2\pi d}{\lambda}\left(\left(n-\dfrac{1+N}{2}\right)\sin\theta\sin\varphi\right)\right)\right) ,
\end{gather}
where $ \theta,\varphi $ are the pitch and azimuth angles relative to RIS, $ f_{mn}(\theta,\varphi) $ stands for the scattering field pattern of the $ mn$-th unit in RIS, $ B_{mn},\beta_{mn} $ stand for the amplitude and phase of the incident wave at the $ mn$-th unit, while $ \lambda $ and $ d $ correspond to the wavelength of electromagnetic wave and the length of spacing between RIS units respectively. The relationship between the antenna gain of RIS in each direction and the scattering pattern is given by \cite{9406940}
\begin{gather}
	\label{model_ris5}
    G_{RIS}(\theta,\varphi)\propto|E(\theta,\varphi)|^{2} .
\end{gather}
 Denote $(\theta_{rx},\varphi_{rx})$ as the direction of the UE. Then we can obtain the relationship between the receiving power $ P_r $ at UE and the transmitting power at BS as:
\begin{equation}
	\label{model_ris_pr}
	P_r = \sum_{k=1}^{K}P_kG_tG_{RIS}(\theta_{rx},\varphi_{rx})G_r\left(\frac{\lambda}{4\pi D}\right)^2 ,
\end{equation}
where $ G_t $ denotes the gain of transmitting antenna, $ G_r $ denotes the gain of receiving antenna, and $ D $ stands for the distance travelled by electromagnetic wave. In addition, $ \left(\lambda/4\pi D\right)^2 $ is the free space path loss (FSPL) of electromagnetic wave.
Typically, $ A_{mn}$ is a constant, and thus we only need to consider the phase modulation of the incident wave. Let us define $ \bm{h}_{ris,\alpha}^{optm} $ as the vector consisting of the optimal modulated phase of RIS units, i.e., $ \bm{h}_{ris,\alpha}^{optm}=[\alpha_{11}^{optm},\cdots,\alpha_{1N}^{optm},\cdots,\alpha_{MN}^{optm}] $. Then \eqref{model_ris3} can be rewritten as
\begin{equation}
	\label{model_ris6}
	\bm{h}_{ris,\alpha}^{optm} =\mathop{\arg\max}\limits_{\alpha_{mn}}\ P_r .
\end{equation}
  According to \eqref{model_ris5}, \eqref{model_ris_pr} and \eqref{model_ris6}, equation \eqref{model_ris3} eventually translates into designing a proper modulated phase $ \alpha_{mn} $ to maximize $ E(\theta_{rx},\varphi_{rx}) $. Then $ \bm{h}_{ris,\alpha}^{optm} $ can be expressed as 
\begin{equation}
	\label{model_ris7}
	\bm{h}_{ris,\alpha}^{optm} =\mathop{\arg\max}\limits_{\alpha_{mn}}\ E(\theta_{rx},\varphi_{rx}) .
\end{equation}
According to (4) and \eqref{model_ris7}, we can calculate the optimal modulation phase as
\begin{equation}
	\label{model_ris8}
	\begin{aligned}
	{\alpha}_{mn}^{\theta_{rx},\varphi_{rx}} =-\dfrac{2\pi d}{\lambda}\left(\left(\dfrac{M+1}{2}-m\right)\cos\theta_{rx} \right) \\
	-\dfrac{2\pi d}{\lambda}\left(\left(n-\dfrac{1+N}{2}\right)\sin\theta_{rx}\sin\varphi_{rx}\right) 
	-\beta_{mn} .
	\end{aligned}
\end{equation}

 Traditionally, we can take the exhaustive search method or beam sweeping method to obtain $(\theta_{rx},\varphi_{rx})$. However, beam sweeping scheme would incur a huge training overhead, which greatly affects the quality of the communication. We next introduce a novel method of using visual information to assist RIS in obtaining $(\theta_{rx},\varphi_{rx})$, which can greatly save the beam training overhead.

\section{Vision-Based Beam Tracking Scheme}
\begin{figure}[!t]
	\centering
	\includegraphics[width=3.5in]{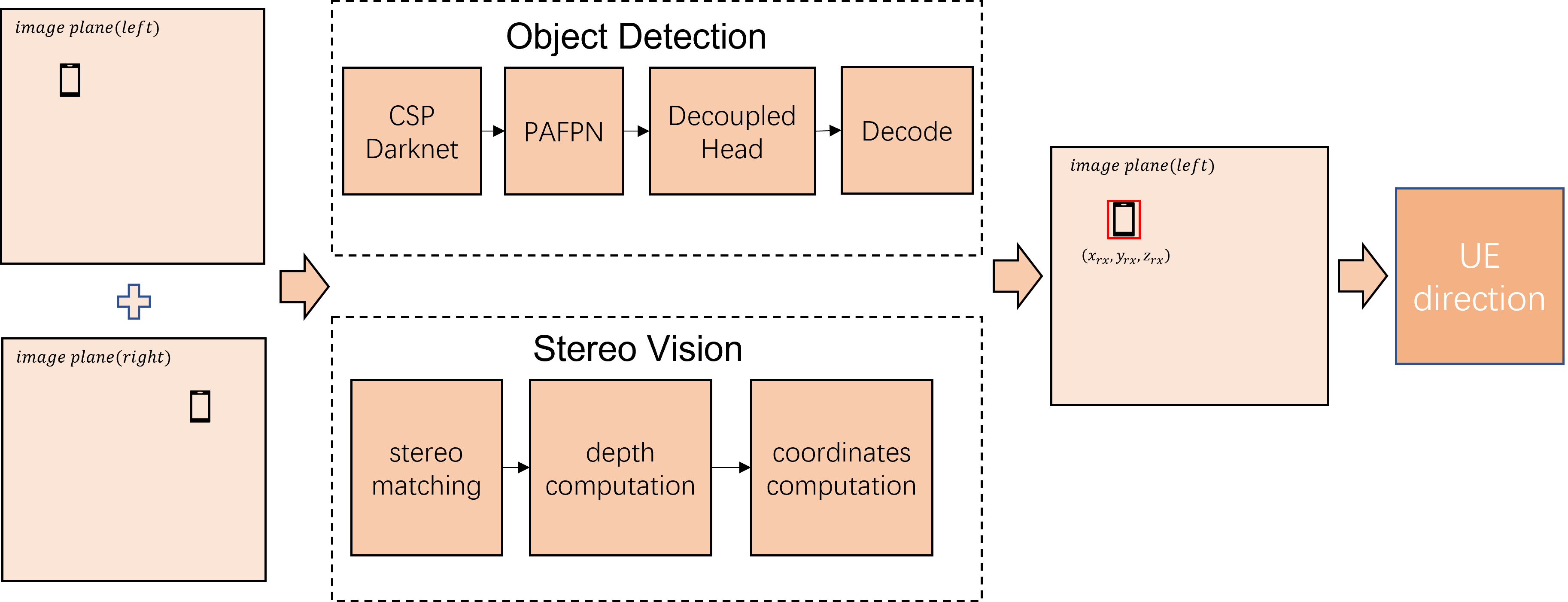}
	\caption{Implementation framework of vision-based beam tracking scheme. Object detection and stereo vision are combined to obtain the UE's 3D coordinates relative to the RIS and then the UE direction can be calculated.}
	\label{fig_2}
\end{figure}

 In the proposed beam tracking scheme, we use a binocular camera to obtain visual information and calculate the UE's direction with advanced object detection algorithm as well as the stereo vision algorithm. The object detection algorithm can provide the UE's 2D coordinates in the image plane, and the stereo vision algorithm can calculate the UE's 3D coordinates with respect to RIS. Based on the 3D coordinates, the direction of UE can be calculated. The implementation framework of the vision-based beam tracking scheme is shown in the Fig.~\ref {fig_2}.

 We choose the latest version of the YOLO (You Only Look Once) algorithm, YOLOX\cite{ge2021yolox}, to realize the object detection. YOLOX is an improved version of YOLOv3, mainly in three aspects. Firstly, the classification task and the regression task are decoupled and implemented separately in one head using two branching networks such that the conflict problem between classification and regression is well resolved. Secondly, the anchor-free scheme is adopted to reduce the number of the network parameters and speed up the prediction. Thirdly, two data enhancement methods, Mosaic\cite{mosaic} and Mixup\cite{mixup}, and the label assignment method SimOTA\cite{ge2021yolox} are introduced to improve the training speed of the network.

In the proposed beam tracking scheme, after YOLOX obtains the environment image from camera, the backbone network CSPDarknet\cite{yolov3} performs feature extraction on the image. The extracted multi-scale features are then fed into the feature pyramid network (FPN) and path aggregation network (PAN) for feature fusion and enhancement. Then three decoupled heads are used for regression and classification, and the output is decoded to obtain the prediction information $ \{x_c,y_c,w,h\} $ of the UE in the image. Note that $ x_c,y_c $ represent the 2D coordinates of the center point of the prediction bounding box while $ w,h $ denote the width and height of the prediction bounding box.

\begin{figure}[!t]
	\centering
	\includegraphics[width=2.5in]{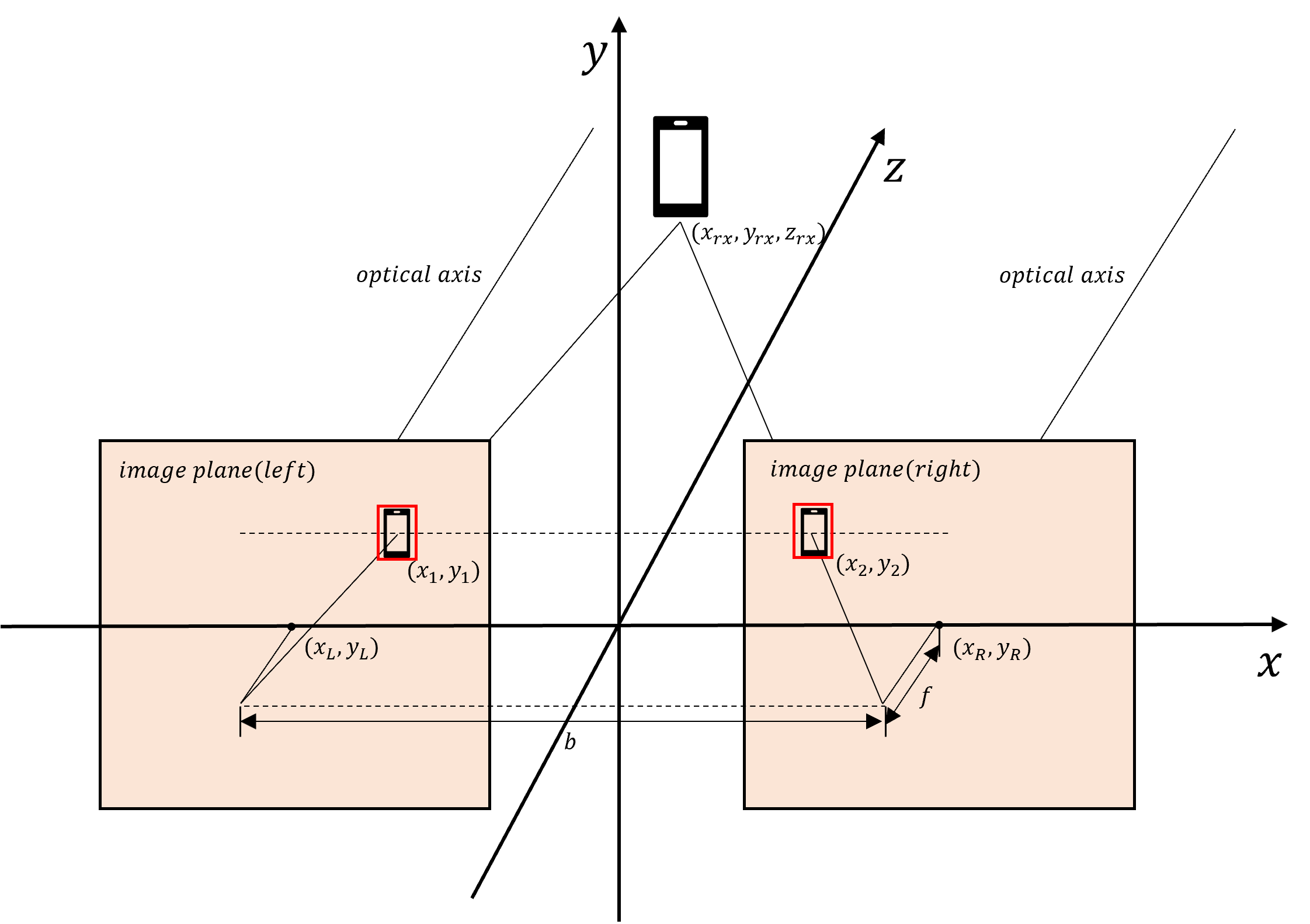}
	\caption{Depth information computation model diagram of the binocular camera. Based on the spacing $ b $ between the two cameras, the focal length $ f $ and the vision disparity of the two images, the depth $z_{rx} $ of the UE can be computed.}
	\label{fig_3}
\end{figure}

However, the object detection algorithm can only provide the 2D coordinates of the UE in the image, while to calculate the 3D coordinates in the real world we have to rely on binocular stereo vision algorithm. In the binocular stereo vision algorithm, firstly stereo matching algorithm is used to obtain the vision disparity between the two images \cite{stereomatch1,stereomatch2}. Then the depth information can be calculated based on the vision disparity. The depth information computation model diagram of the binocular camera is shown in Fig.~\ref {fig_3}, where $ (x_L,y_L),(x_R,y_R)$ denote the coordinates of the left and the right image centroids, $ (x_1,y_1),(x_2,y_2) $ denote the 2D coordinates of the UE in two images, $ (x_{rx},y_{rx},z_{rx})$ denote the 3D coordinates of the UE with respect to RIS, and $ f,b$ denote the focal length and the spacing between the two cameras. Typically, the vision disparity $ d$ is defined as
\begin{equation}
	\label{vsion_solu1}
	d = x_1-x_L+x_R-x_2 .
\end{equation}
Based on the binocular geometry \cite{stereovision}, we can derive the depth $ z_{rx}$ of the UE as
\begin{equation}
	\label{vsion_solu2}
	z_{rx} = \dfrac{fb}{d} .
\end{equation}

Combining the depth $ z_{rx}$, the intrinsic matrix of the camera and the 2D coordinates $ (x_c,y_c) $ output by YOLOX, we can calculate the 3D coordinates $ (x_{rx},y_{rx},z_{rx})$ of the UE with respect to RIS and then obtain the the pitch angle $\theta_{rx}$ and azimuth angle $\varphi_{rx}$ of the UE with respect to RIS. Given $\theta_{rx}\in[0,\pi]$ and $ \varphi_{rx}\in[-\frac{\pi}{2},\frac{\pi}{2}] $, the direction $(\theta_{rx},\varphi_{rx})$ of the UE can be calculated as:
\begin{gather}
	\label{vsion_solu3}
    \begin{align} 
		&\theta_{rx}=\left\{
	\begin{array}{rcl}
		\arctan\left(\dfrac{\sqrt{x_{rx}^2+z_{rx}^2}}{y_{rx}}\right) & & {y_{rx}>0,}\\
		\pi+\arctan\left(\dfrac{\sqrt{x_{rx}^2+z_{rx}^2}}{y_{rx}}\right) & & {y_{rx}<0 ,}
	\end{array} \right.                                                       
    \\
	&\varphi_{rx}=\arctan\left(\dfrac{x_{rx}}{z_{rx}}\right) .
   \end{align} 
\end{gather}
According to \eqref{model_ris8}, (12) and (13), we can calculate and adjust the optimal reflection coefficients of RIS to achieve beam tracking.

\section{System Design}
\begin{figure*}[!t]
	\centering
	\subfloat[]{\includegraphics[width=3.5in]{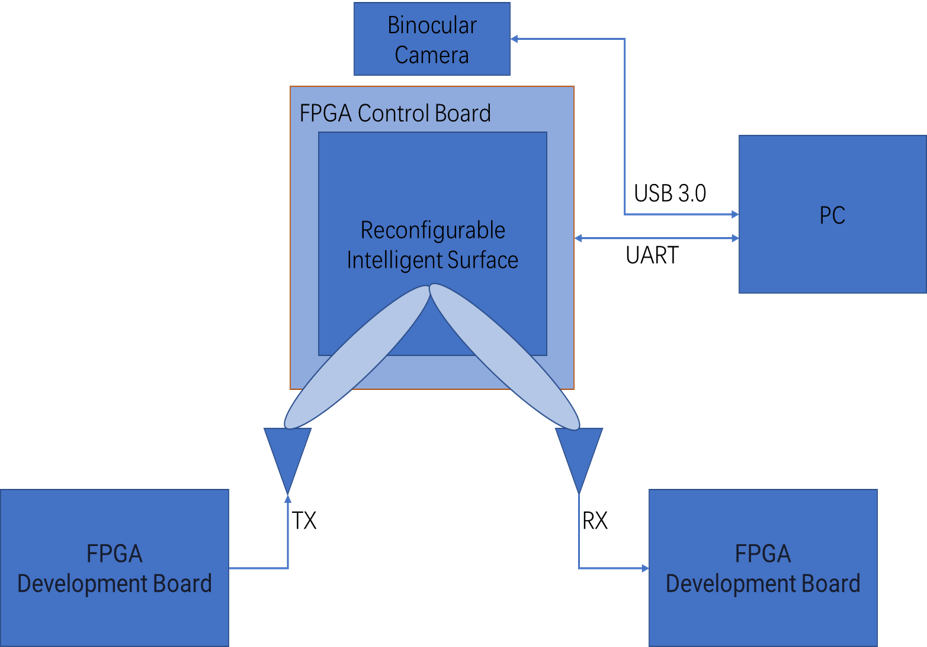}%
		\label{systemarch}}
	\hfil
	\subfloat[]{\includegraphics[width=3.5in]{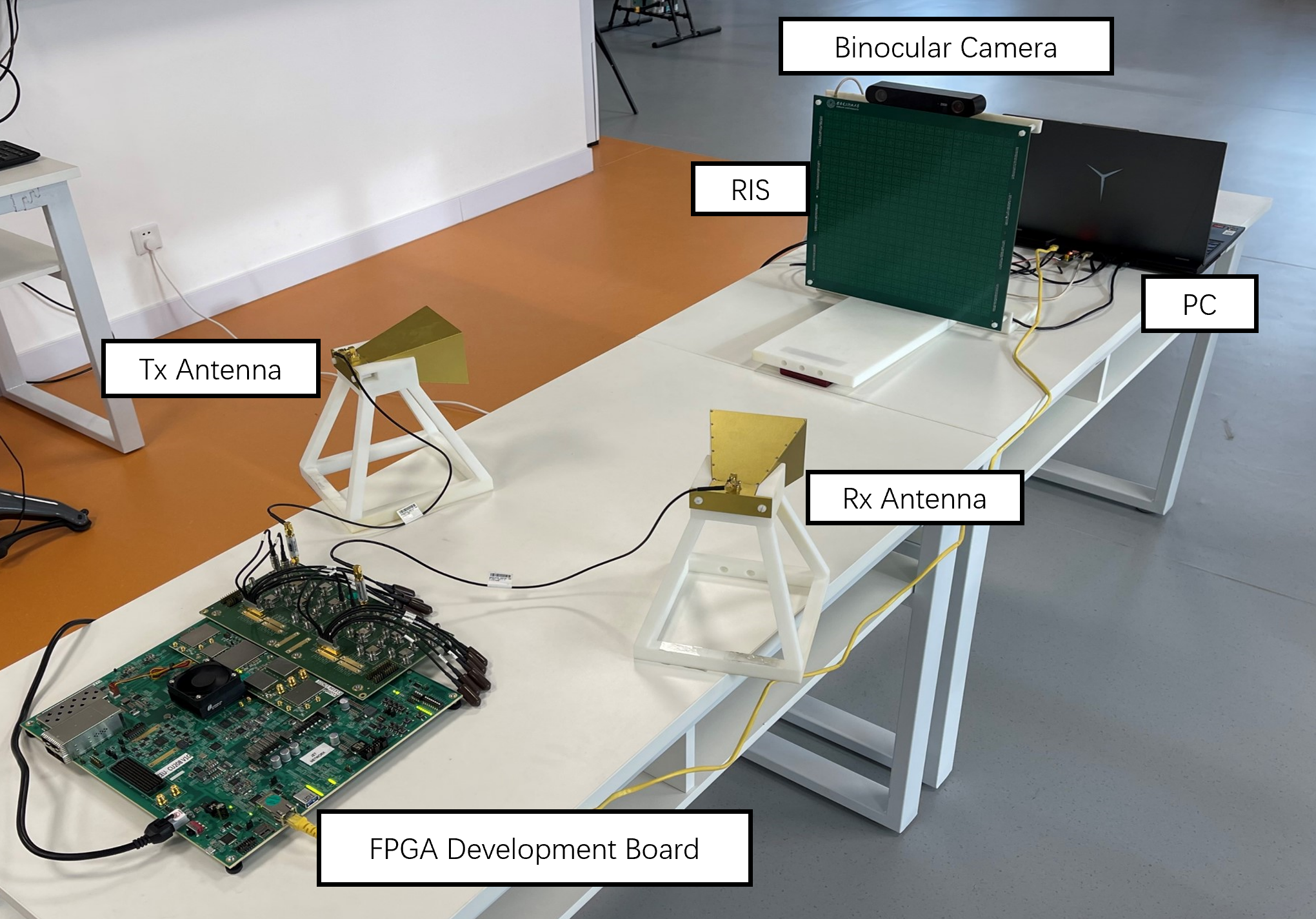}%
		\label{testscene_note}}
	\caption{Architecture diagram of the prototype system. The PC gets the environmental information obtained by the cameras through the USB interface, and then the RIS gets the UE's direction from the PC through the serial port and adjusts the reflection coefficients accordingly. (a) Theoretical block diagram of the prototype system. (b) Physical diagram of the prototype system. }
	\label{systemarch_fig}
\end{figure*}

The architecture of the proposed prototype system is shown in the Fig.~\ref {systemarch_fig}, where Fig.~5(a) presents the theoretical block diagram and Fig.~5(b) presents the physical diagram. In this section, we present the system design in four aspects, including vision module, RIS and codebook, control board, and peer-to-peer communication system.

\subsection{Vision Module Design}
The binocular camera we adopted is the ZED II camera provided by Stereolabs. ZED II camera takes deep learning-based stereo matching algorithm to calculate the vision disparity and then calculates the 3D coordinates of the pixels based on the vision disparity. In addition, Stereolabs provides a very comprehensive API to obtain depth information. Therefore, after YOLOX outputs the 2D coordinates of the UE in the image plane, we only need to call the corresponding interface function of the ZED II camera to obtain the 3D coordinates of the UE with respect to RIS.

The YOLOX algorithm and the stereo matching algorithm both are running on the personal computer (PC). ZED II camera sends the environmental photos to PC through the USB interface, and then PC sends the calculated receiver direction to RIS control board through the serial port. It is tested that the whole process from acquiring photos to outputting UE's direction takes about 85 ms. Moreover, to ensure detection accuracy, the YOLOX-DarkNet53 model is used in the designed vision module. However, a lighter version can be used to obtain faster inference, e.g., YOLOX-Tiny.
\subsection{RIS and Codebook Design}
\begin{figure}[!t]
	\centering
	\includegraphics[width=3.5in]{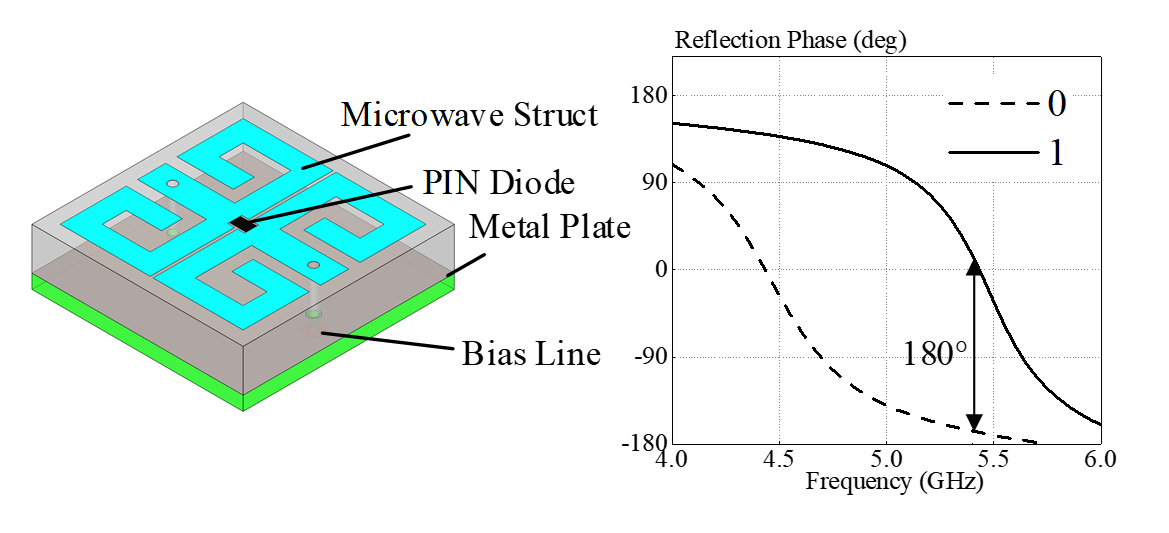}
	\caption{Structure diagram and modulation phase response diagram of RIS unit. The diagram on the left shows the unit consisting of the top microwave structure, a metal plate, a bias line, and a PIN diode. The diagram on the right shows how the modulation phase of the RIS unit varies with the operating frequency, where the phase difference between the ``0'' and ``1'' states is 180 degree at 5.4 GHz. }
	\label{risunitdesign}
\end{figure}

The RIS in Fig. \ref {systemarch_fig} contains 400 (20$\times$20) units with a quarter wavelength spacing between the individual units. Meanwhile, the units are square structure with quarter wavelength sides, as shown in Fig.~\ref {risunitdesign}. Each unit consists of a microwave structure on the top layer, a metal plate, a bias line and a PIN diode \cite{9643997}. The bias voltage across the PIN diode can be changed to control the PIN diode states (forward bias state or reverse bias state). Here we denote ``1'' state as forward bias state and ``0'' as the reverse bias state. For different operating frequencies, the unit has different modulation phases. The modulation phase difference between ``0'' state and ``1'' state should be 180 degree. However, PIN diodes with different batches have different equivalent circuit parameters, and thus the most suitable operating frequency of the unit is decided based on the test results. The modulation phase of the RIS unit varies as a function of operating frequency as shown in Fig.~\ref {risunitdesign}. It is seen that the optimal operating frequency is 5.4 GHz.
Generally, the specific modulation phase values corresponding to the two states are not determined. For the convenience of the codebook design, we assume that the modulation phase of the unit controlled by the PIN diode in ``1'' state is $-\pi/2$, and the modulation phase of the unit controlled by the PIN diode in the ``0'' state is $\pi/2$.

According to \eqref{model_ris8}, in order to calculate the optimal modulation phase ${\alpha}_{mn}^{\theta_{rx},\varphi_{rx}}$, we should first obtain the initial phase $ \beta_{mn} $ of the incident wave. The way to calculate the phase $ \beta_{mn} $ is not fixed and can be different for two cases: (\uppercase\expandafter{\romannumeral1}) RIS works in near-field as a passive array antenna of the BS; (\uppercase\expandafter{\romannumeral2}) RIS works in far-field to assist the communication between the BS and the UE.

In case \uppercase\expandafter{\romannumeral1}, the distance between the transmitting antenna and the RIS is not very different from the maximum aperture of the RIS, and thus we need to consider the distance between the transmitting antenna and each unit precisely. We assume that the center of the transmitting antenna is facing the center of the RIS and denote the vertical distance between the transmitting antenna and the RIS as $ d_{feed} $. The incident wave phase $ \beta_{mn}^{I} $  of  $ mn$-th unit can be calculated as

\begin{gather}
	\label{design_ris2}
     {\beta}_{mn}^{I} =\dfrac{2\pi }{\lambda} d_{feed}- \notag \\
        \dfrac{2\pi }{\lambda}\sqrt{d^2\left(\dfrac{M+1}{2}-m\right)^2 +d^2\left(n-\dfrac{1+N}{2}\right)^2+d_{feed}^2}  \  .
\end{gather}

In case \uppercase\expandafter{\romannumeral2}, the incident electromagnetic waves can be approximated as plane waves with respect to the RIS. We assume that the pitch and azimuth angles of the transmitting antenna with respect to the RIS are $\theta_{tx}$ and $\varphi_{tx}$, respectively. Then we can obtain the incident phase $ \beta_{mn}^{II} $ of  $ mn_{th} $ unit as
\begin{equation}
	\label{design_ris3}
	\begin{aligned}
		{\beta}_{mn}^{II} =\dfrac{2\pi d}{\lambda}&\left(\left(\dfrac{M+1}{2}-m\right)\cos\theta_{tx}\right)  \\
		&+\dfrac{2\pi d}{\lambda}\left(\left(n-\dfrac{1+N}{2}\right)\sin\theta_{tx}\sin\varphi_{tx}\right).
	\end{aligned}
\end{equation}

\begin{figure}[!t]
	\centering
	\includegraphics[width=3.5in]{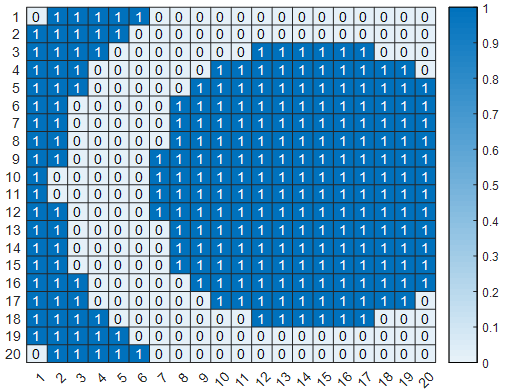}
	\caption{the RIS code word in case I when $\theta_{rx}= 90^{\circ},\varphi_{rx}= 10^{\circ}$. The 0 in the figure represents the modulation phase of $\frac{\pi}{2}$, while the 1 represents $-\frac{\pi}{2}$. }
	\label{visionriscode_fig}
\end{figure}

The optimal modulation phase ${\alpha}_{mn}^{\theta_{rx},\varphi_{rx}}$ can be obtained by substituting (14) or \eqref{design_ris3} into \eqref{model_ris8}. However, the individual units can only provide two different phases $\left(\pi/2 \ \text{or}\ -\pi/2\right)$, and thus we need to quantize ${\alpha}_{mn}^{\theta_{rx},\varphi_{rx}}$ by one bit. The specific scheme of the quantization is
\begin{equation}
	\label{design_ris1}
	{\alpha}_{mn,quan}^{\theta_{rx},\varphi_{rx}}=\left\{
	\begin{array}{rcl}
		\frac{\pi}{2} & & {{\alpha}_{mn}^{\theta_{rx},\varphi_{rx}} \in [0,\pi) ,}\\
		-\frac{\pi}{2} & & {{\alpha}_{mn}^{\theta_{rx},\varphi_{rx}} \in [-\pi,0) .}
	\end{array} \right.
\end{equation}
According to \eqref{model_ris8}, (14), \eqref{design_ris3}, and \eqref{design_ris1}, we can arbitrarily set the desired outgoing direction $(\theta_{rx},\varphi_{rx})$ to get the corresponding codebook, e.g., under case \uppercase\expandafter{\romannumeral1} we can calculate the RIS codeword when the desired pitch angle $\theta_{rx}$ is 90 degree and the azimuth angle $\varphi_{rx}$ is 10 degree, as shown in Fig.~\ref {visionriscode_fig}. The ``0'' in the Fig.~\ref {visionriscode_fig} represents the PIN diode reverse bias and the modulation phase of the controlled unit is $\pi/2$. The ``1'' in the Fig.~\ref {visionriscode_fig} represents the PIN diode forward bias and the modulation phase of the controlled unit is $-\pi/2$. Based on the proposed codeword design principle, the codebook can be pre-calculated and solidified.

\subsection{Control Board Design}
 In order to meet the real-time beam tracking in the mobile scene, the vision-aided RIS needs to have a fairly fast beam switching speed, which imposes high requirements on the code switching speed. Therefore, we cascaded two low-cost, high-I/O-density FPGA chips (Intel Cyclone IV EP4CE15F23C8N) in the control board, which can easily run at up to 200 MHz system clock frequency and control each PIN diode individually. The codeword switching speed is much faster than the general shift register scheme \cite{trichopoulos2021design}.

\begin{figure}[!t]
	\centering
	\includegraphics[width=3in]{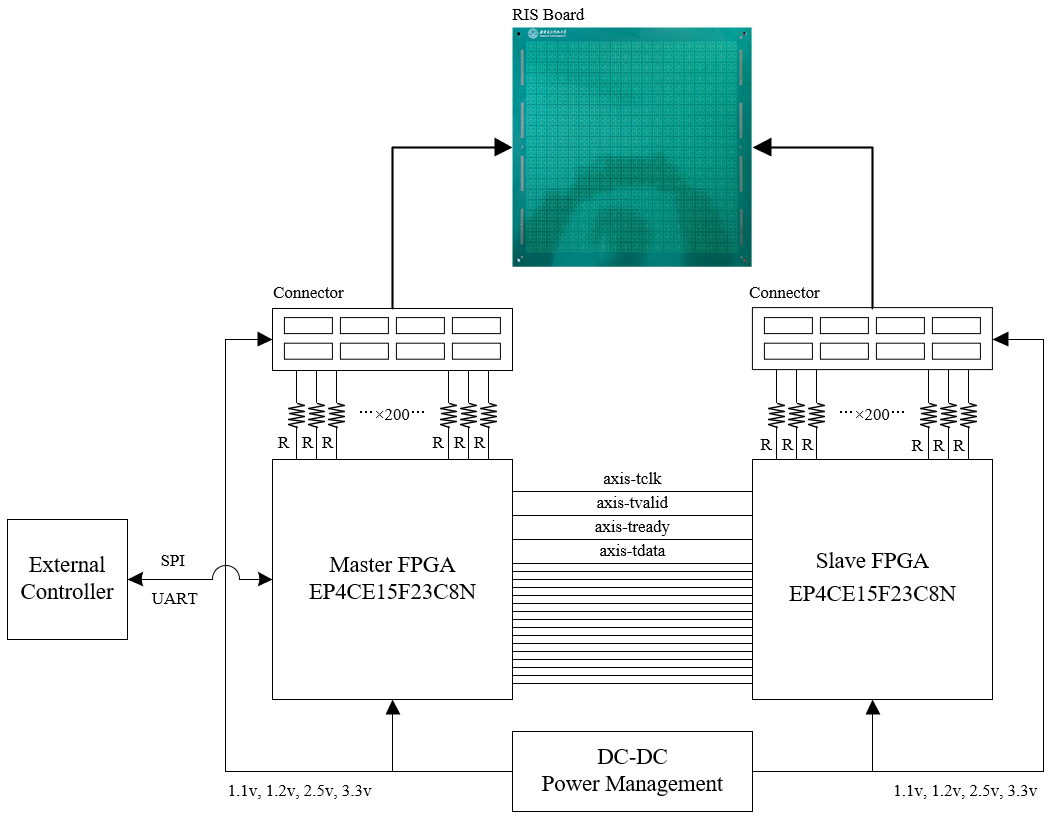}
	\caption{Schematic diagram of RIS High-Speed control board. The master FPGA chip can receive codewords or external instructions through the external interface and can control the slave FPGA or transmit codewords through the AXI-stream-4 bus. The control board can control the state of each RIS unit independently through the I/O pins of the FPGAs.}
	\label{visionriscontrolboarddesign.fig}
\end{figure}

 Fig.~\ref {visionriscontrolboarddesign.fig} shows the schematic diagram of the RIS control circuit, including a master FPGA controller, a slave FPGA controller, an external communication interface, bias circuits and power management module. The master FPGA is responsible for controlling the first 10 rows (200 in total) of PIN diodes and external communication, and the slave FPGA is responsible for controlling the other 200 PIN diodes. The master FPGA and the slave FPGA are connected on the printed circuit board through a parallel 16-bit AXI-Stream-4 bus, where the signal line consists of 16-bit-tdata, 1-bit-tready, 1-bit-tvalid, and 1-bit-tclk. The bus follows the AXI-Stream-4 protocol for communication. The reference clock rate of 100 MHz allows for high speed data transfer from master to slave FPGA chips, which can reach a peak speed of 1.6 Gbps.

\begin{figure}[!t]
	\centering
	\includegraphics[width=3in]{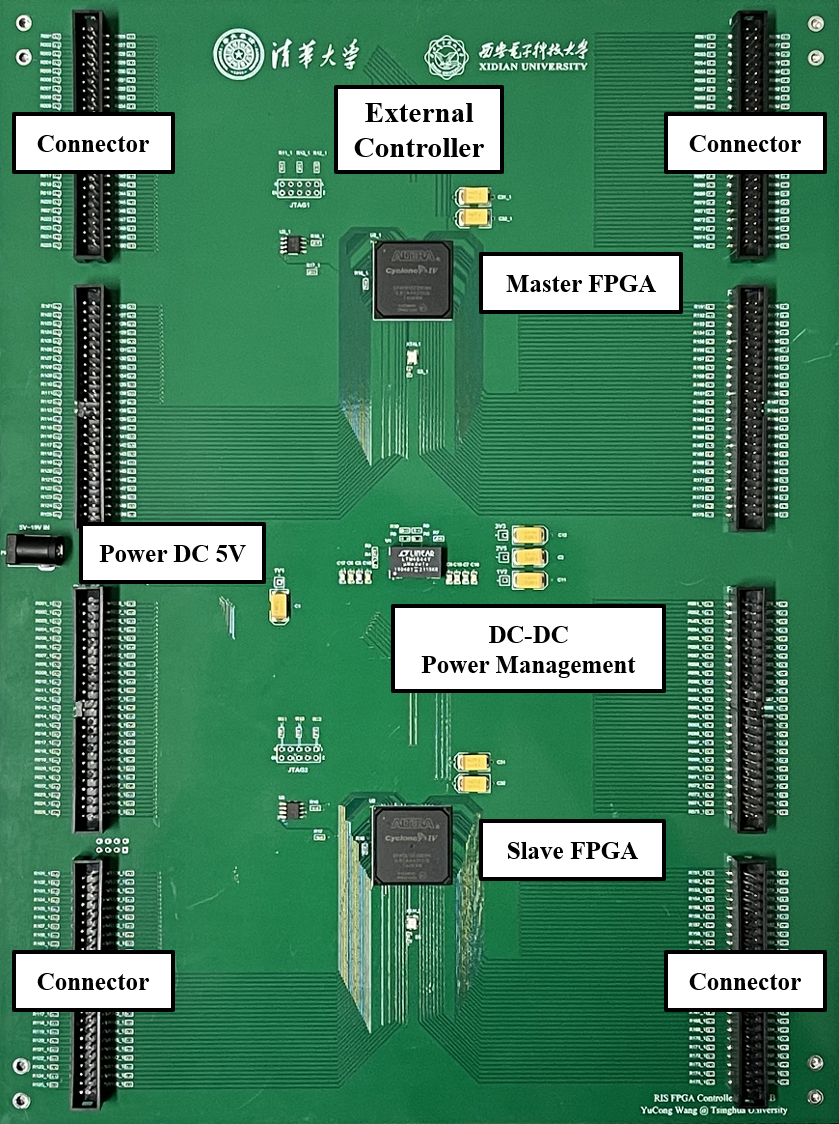}
	\caption{RIS control board physical diagram. Each part of the diagram corresponds the schematic. The control board can be directly connected to the RIS board through eight 2.54 mm connectors.}
	\label{riscontrolboard.fig}
\end{figure}

 In the designed bias circuits, each I/O of the FPGA chips is connected in series with PIN diodes, current limiting resistors and a 1.1 v voltage source. Each I/O can output 3.3 v or 0 v voltage (1 or 0) to provide each PIN diode with 1.5 v forward bias or 1.1 v reverse bias and to limit current to 5 mA. The forward bias state of the PIN diode corresponds to the ``1'' in the codeword, and the reverse bias state corresponds to the ``0''. In addition, the power management part is designed with an integrated 4-channel DC-DC chip LTM4644IY, which generates the 1.2 v, 2.5 v, 3.3 v power rails required by the FPGA chips, and the 1.1 v reverse bias voltage required by the PIN diode. The 400 control I/O signals are connected to the RIS front panel via eight 2.54 mm connectors. Fig.~\ref {riscontrolboard.fig} shows the finished control board. Since the current passing through each PIN diode is very small, the total power consumption of the RIS including the bias circuit and FPGA is less than 0.5 W.

 In order to enhance the robustness and universality of the prototype system, the control board has three working modes:
\subsubsection*{\bf Index control mode}
 In this mode, the pre-calculated codebook needs to be downloaded to the Flash memory of the FPGA in advance. The PC selects the appropriate codeword to be switched and transmits the corresponding index number to the master FPGA through the external communication interface (serial port, SPI). Then the master FPGA sends the index number to the slave FPGA through the parallel AXI-Stream-4 bus.
\subsubsection*{\bf Dynamic codebook mode}
In this mode, the codewords are calculated on the PC in real time. Then the codeword stream should be inputted through the external communication interface, and the master FPGA sends codeword to the slave FPGA through the parallel AXI-Stream-4 bus.
\subsubsection*{\bf Codebook download mode}
After power-on, the FPGA chips rewrite the preset codebook through the external communication interface, and then the master FPGA sends the half of the codebook to the slave FPGA through the parallel AXI-Stream-4 bus.


\begin{figure}[!t]
	\centering
	\includegraphics[width=3in]{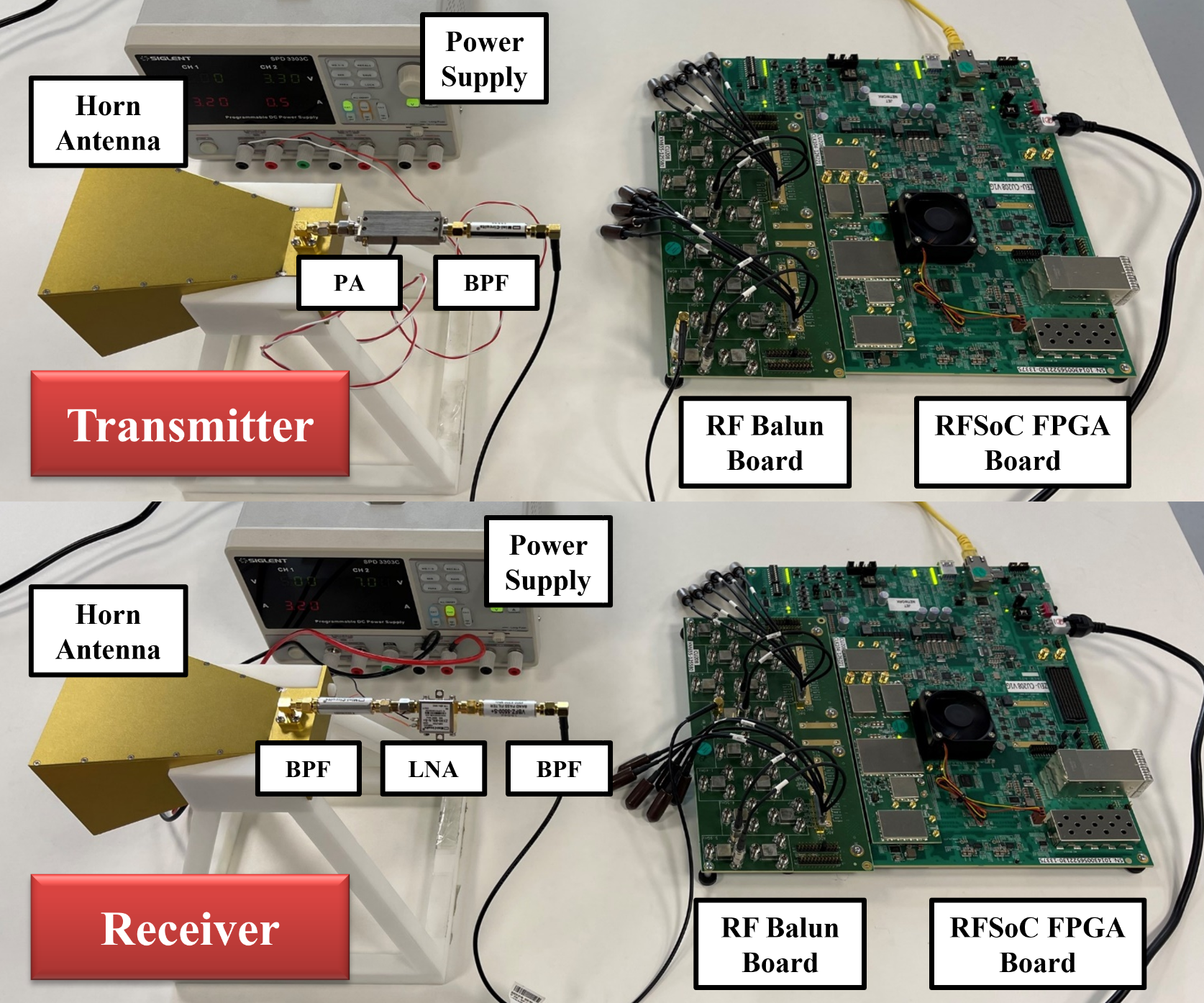}
	\caption{Processing link diagram for peer-to-peer communication system.}
	\label{visionrisphyrx.fig}
\end{figure}
\subsection{Peer-to-peer Communication System Design}
 To validate the effectiveness of the vision-aided RIS, we design a peer-to-peer communication system to simulate the communication between the BS and the UE, as shown in Fig.~\ref {visionrisphyrx.fig}. The RFSoC development board, the transmitting RF chain, and the transmittting horn antenna are the main components of the transmitter, which emulates the BS. The RFSoC development board, receiving RF chain, and receiving horn antenna are the main components of the receiver, which emulates the UE. In the transmitter, the business data is transmitted from PC to the ARM processor of the RFSoC chip through User Datagram Protocol (UDP), which runs the LwIP protocol stack to parse the UDP packets. In addition, there is a simple MAC program to package the data into physical layer frames, and then ARM processor sends the data to programmable logic part via XDMA and AXI-S bus. Modulation and demodulation Verilog programs is running in the programmable logic part to process digital baseband signals in real time. Meanwhile, the receiver demodulates the bit stream and transmits it to the host via UDP. The digital baseband can also upload the intermediate information generated in the demodulation to the PC and display the intermediate information, such as constellation diagram, channel status information, etc.

\begin{figure}[t]
	\centering
	\includegraphics[width=3.5in]{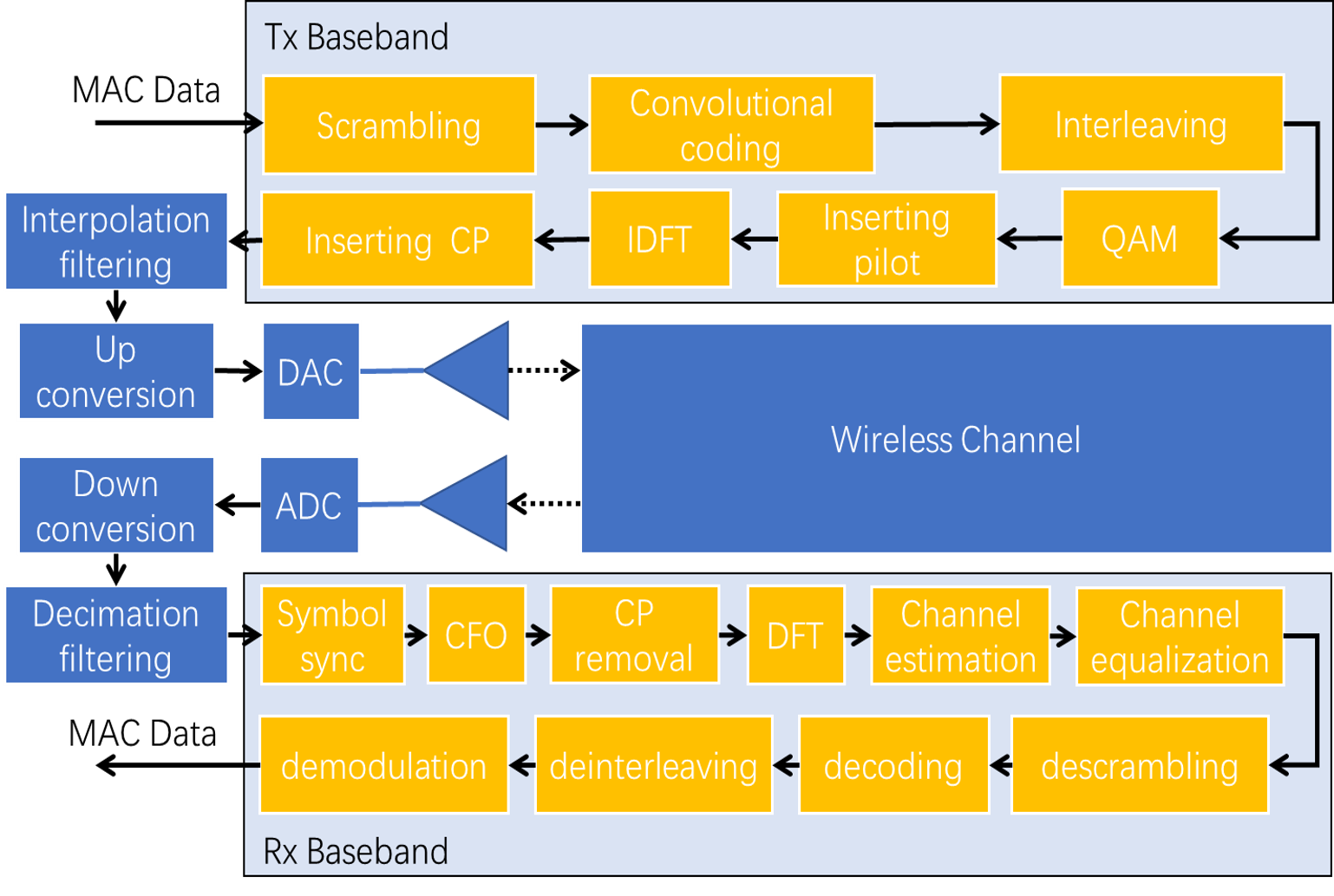}
	\caption{Signal processing flow block diagram for peer-to-peer communication system. Except for the analog filtering and amplification process, the digital processing section (including digital baseband and up/down conversion) is shown in this figure. }
	\label{baseband.fig}       
\end{figure}

The baseband algorithm of the system is implemented strictly with reference to the 802.11n standard of WiFi, with a bandwidth of 40MHz. Based on this standard, the whole baseband contains 52 data subcarriers and 4 pilot subcarriers with a subcarrier spacing of 625 kHz. The baseband algorithm is divided into two parts, i.e., the transmitter and the receiver, to be implemented separately. The baseband algorithm of the transmitter mainly includes seven steps: scrambling, convolutional coding, interleaving, quadrature amplitude modulation (QAM), insertion of pilot, inverse discrete fourier transform (IDFT) and insertion of cyclic prefix (CP). At the receiver side, the wireless signal is demodulated by the baseband algorithm to obtain the transmitted data after down-conversion and decimation filtering. The receiver’s baseband algorithm consists of ten steps: symbol synchronization, carrier frequency offset (CFO) estimation and compensation, CP removal, DFT, channel estimation, channel equalization, symbol demodulation, deinterleaving, decoding, and descrambling. The signal processing flow of the peer-to-peer communication system is shown in Fig.~\ref {baseband.fig}.

Next, we will give a detailed description of the processing of the RF link. After the 40 MHz baseband I/Q signal is generated by the digital transmit baseband, the baseband signal is interpolated to the sampling rate of 4.8 GHz, and then is mixed with a 48-bit digital oscillator (NCO). The baseband signal is placed at 5.4 GHz on the spectrum via digital up-conversion (DUC) and is sent to the RF-DAC. The RF-DAC supports bandpass sampling and is set to the third Nyquist zone mode, which can optimize the signal power and in-band flatness in the third zones.
 Since the signal generated by RF-DAC has image signals and the second harmonics, the transmit RF chain is configured with a 5.2 GHz-5.8 GHz band-pass filter to remove unwanted spurious signals. After filtering, the RF signal passes through the driver amplifier and power amplifier, and is transmitted through the horn antenna. Similarly, in the transmitter, the receiving RF chain is configured with a low-noise amplifier and a band-pass filter. The RF-ADC has a sampling rate of 4.8 GHz and works in the third Nyquist zone. The sampled signal is decimation filtered and digital down-converted (DDC) to obtain the baseband I/Q signal.

After testing, the peak transmission rate of the system can reach 52.4 Mbps on 16QAM modulation, which can transmit 2K high-definition video encoded by H.264 in real time. Additionally, the system can display the demodulation constellation diagram in real time. The parameters of the peer-to-peer communication system are shown in Table~\ref{tab:table1}, which meets the verification requirements of the vision-aided RIS.

\begin{table}[!t]
\caption{Communication System Parameter\label{tab:table1}}
\centering
\begin{tabular}{c c}
\bottomrule
Parameter & Value  \\ \bottomrule
Modulation & BPSK,QPSK,16QAM    \\ \midrule
Bandwidth & 40 MHz    \\ \midrule
Receive Antenna Gain & 7 dBi \\ \midrule
Transmit Antenna Gain & 7 dBi \\ \midrule
Receive Link Gain & 22 dB \\ \midrule
Transmit Link Gain & 30 dB \\ 
\bottomrule
\end{tabular}
\end{table}

\section{Experimental Results}
In this section, we simulate the radiation pattern of the RIS and compare the simulation results with the radiation patterns of the actual test. Then we examine the beam tracking effect of the vision-aided RIS in two classical cases.

\subsection{Radiation Pattern of RIS}

\begin{figure}[!t]
	\centering
	\includegraphics[width=3.5in]{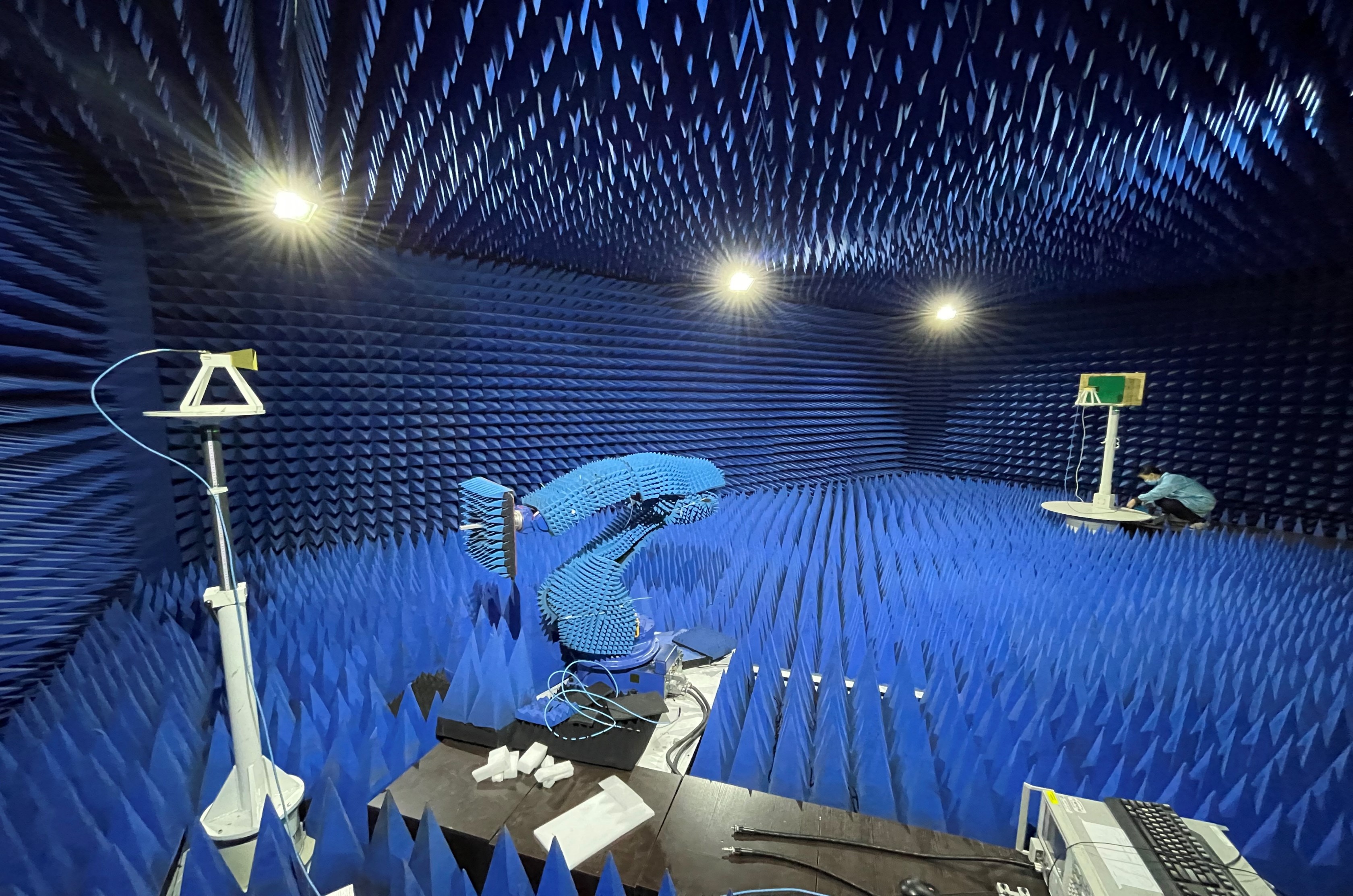}
	\caption{Test scene for practical radiation patterns. The entire test is performed in an anechoic chamber with the RIS on an antenna turntable, and the antenna gains are measured in all directions by a vector network analyzer.}
	\label{testscene.fig}
\end{figure}

Taking the case I where the RIS works in near-field condition as an example, we assume that the transmitting antenna is located at three electromagnetic wave wavelengths directly in front of the RIS. According to \eqref{model_ris8}, (14) and \eqref{design_ris1}, we fix the desired pitch angle $\theta_{rx}$ as $90^{\circ}$ and the desired azimuth angle $\varphi_{rx}$ as $-40^{\circ},-30^{\circ},-20^{\circ},-10^{\circ},0^{\circ},10^{\circ},20^{\circ},30^{\circ},40^{\circ}$, respectively, to get the corresponding codewords. 
Then we use the three-dimensional electromagnetic field simulation software CST to calculate the radiation pattern of the RIS as shown in Fig.~13(a). As can be seen from the figure, the angles of the centers of the main lobes of radiation patterns are consistent with the desired angles, which verifies the correctness of the codebook design principle.
Meanwhile, we test the practical radiation patterns of the RIS in an anechoic chamber free of other electromagnetic wave interference, as shown in Fig.~\ref {testscene.fig}. The test tool is the vector network analyzer and the antenna test turntable system. Based on the codewords calculated in the above radiation pattern simulation, we obtain the practical radiation patterns of multiple reception angles as shown in Fig.~13(b). The results are basically consistent with the simulation results, which is further proof of the correctness and the feasibility of the proposed codebook design principle.

\begin{figure*}[!t]
	\centering
	\subfloat[]{\includegraphics[width=3in]{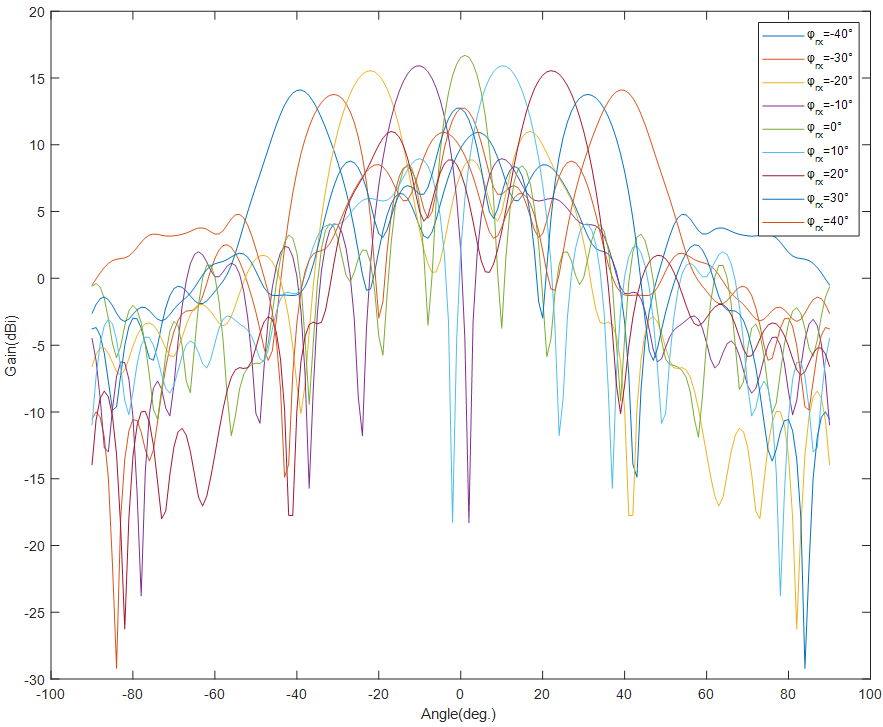}%
		\label{simrispatternall}}
	\hfil
	\subfloat[]{\includegraphics[width=3in]{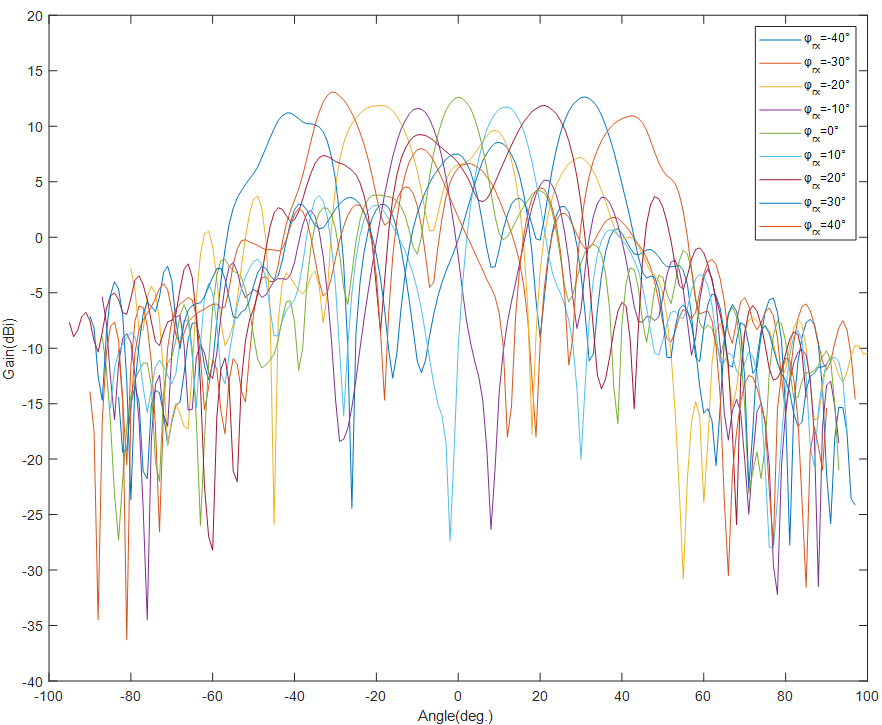}%
		\label{realrispatternall}}
	\caption{Radiation patterns of RIS. The desired pitch angle $\theta_{rx}$ of the reflection beam is fixed as $90^{\circ}$ and the desired azimuth angle $\varphi_{rx}$ is set as $-40^{\circ},-30^{\circ},-20^{\circ},-10^{\circ},0^{\circ},10^{\circ},20^{\circ},30^{\circ},40^{\circ}$, respectively. Nine RIS radiation patterns are simulated or tested. (a) Simulation results of RIS radiation patterns. (b) Practical test results of RIS radiation patterns. }
	\label{rispattern_fig}
\end{figure*}

\subsection{Control Board Test}
\begin{figure}[!t]
	\centering
	\includegraphics[width=3.5in]{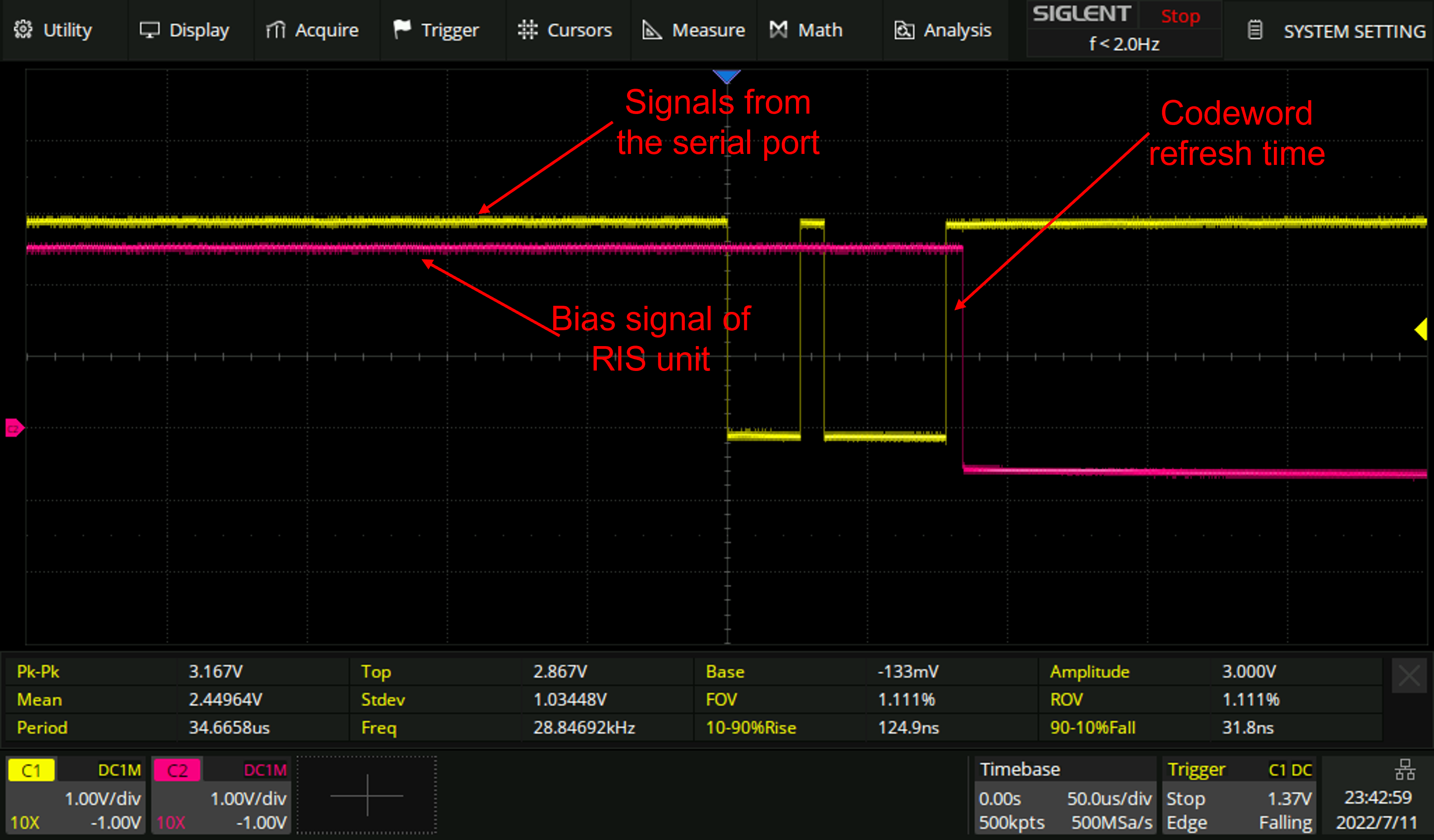}
	\caption{Codeword refresh time test. The first signal connected to the oscilloscope is the control signal sent by the PC through the serial port, and the other signal is the bias signal on the diode.}
	\label{controltest.fig}
\end{figure}

In the proposed prototype system, in order to ensure the codeword refresh speed, the control board uses two FPGA chips to control each diode on the RIS independently. Here we use an oscilloscope to test the codeword switching time to check the efficiency of the control board. The first signal connected to the oscilloscope is the control signal sent by the PC through the serial port, and the other signal is the bias signal on the diode. The test results are shown in Fig.~\ref{controltest.fig}, from which it can be seen that the time from the control signal sent to the completion of the codeword refresh is about 85 us.
\subsection{Test Under the Case \uppercase\expandafter{\romannumeral1}}
\begin{figure}[!t]
	\centering
	\includegraphics[width=3.5in]{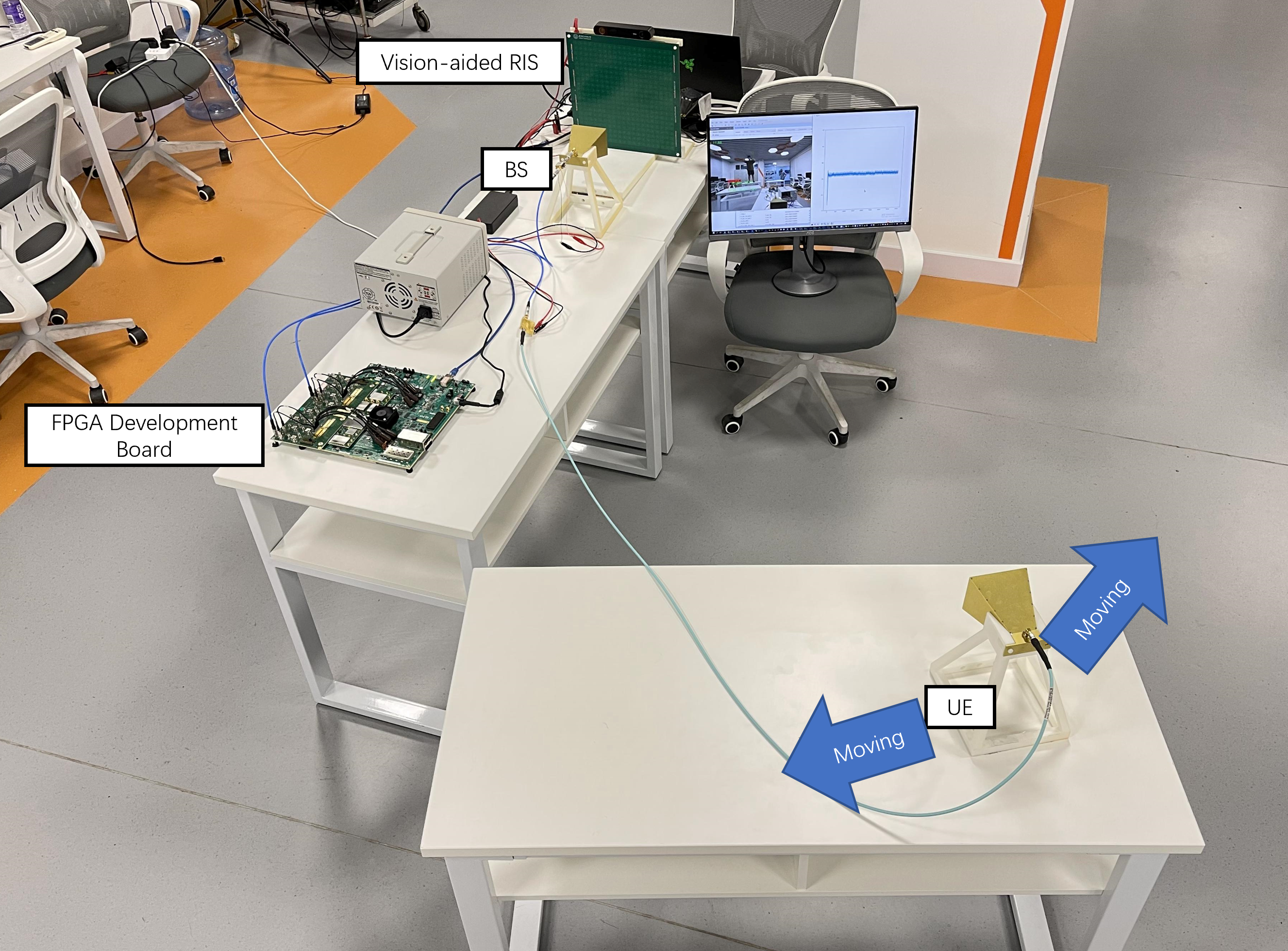}
	\caption{Test scenario layout under case \uppercase\expandafter{\romannumeral1}. We use the transmitting horn antenna located at a distance of 3 electromagnetic wavelengths in front of the RIS as the feeding antenna of the BS and use the receiving horn antenna at a distance of 3 meters from the RIS as the UE.}
	\label{SystemlayoutScenario1.fig}       
\end{figure}

\begin{figure}[!t]
	\centering
	\includegraphics[width=3.5in]{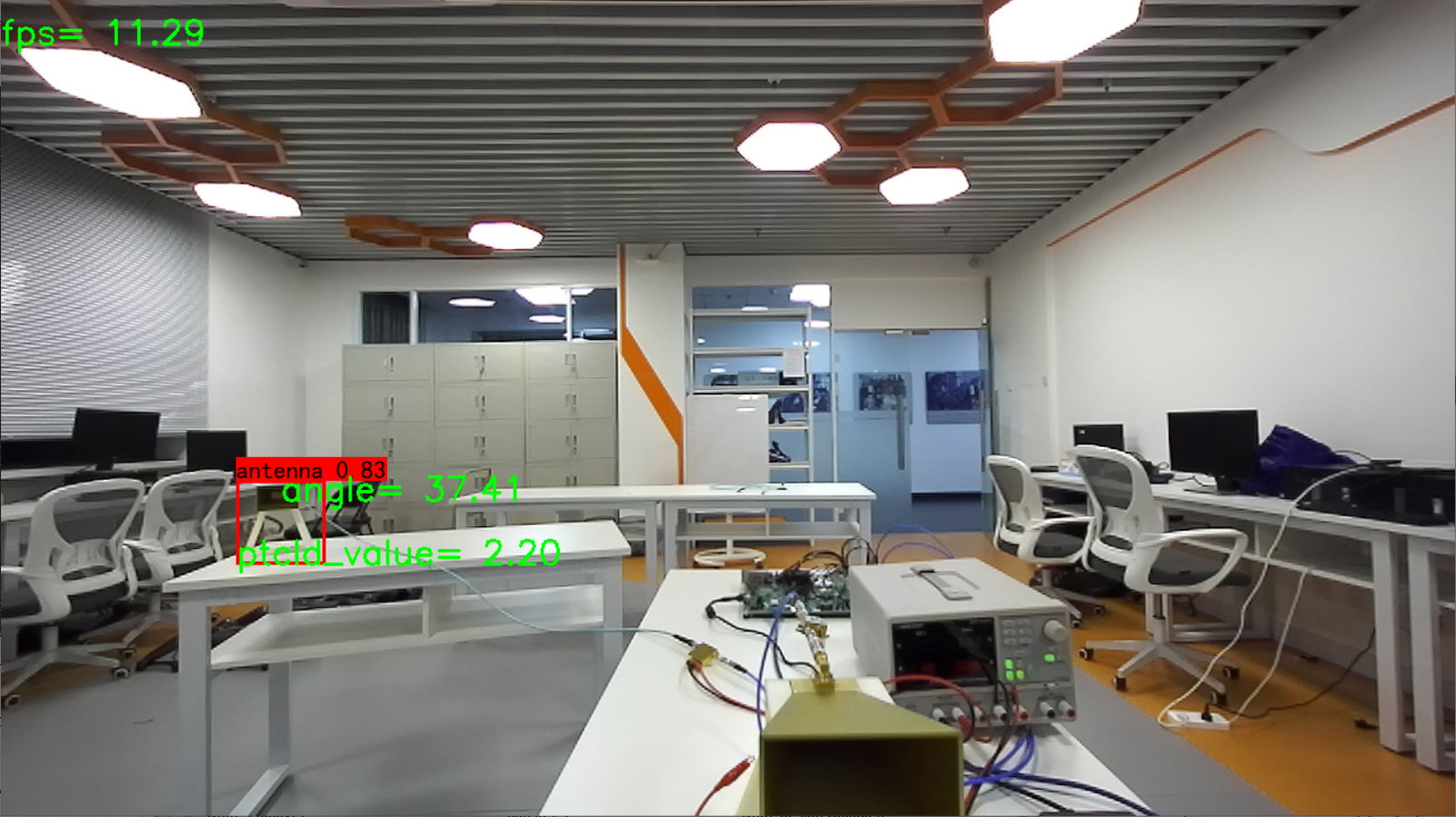}
	\caption{The UE's prediction bounding box and the UE's direction output by vision algorithm at a certain moment.}
	\label{positionofthereceiver.fig}       
\end{figure}

 In case \uppercase\expandafter{\romannumeral1}, the RIS works as a passive array antenna at BS to achieve beamforming. In the test scenario, we use the transmitting horn antenna located at a distance of 3 electromagnetic wavelengths in front of the RIS as the feeding antenna of the BS and use the receiving horn antenna at a distance of 2.2 meters from the RIS as the UE. The layout of the test scenario is shown in Fig.~\ref {SystemlayoutScenario1.fig}. The task of the vision-aided RIS is to accurately reflect the electromagnetic waves emitted by the feeding antenna to the horn antenna at the receiver side based on the visual information. Under case \uppercase\expandafter{\romannumeral1}, the RIS works in near-field condition, and thus we can calculate the RIS codebook in advance according to \eqref{model_ris8}, (14) and \eqref{design_ris1}.
As for the vision-based beam tracking scheme, we use object detection and stereo vision technology to obtain the UE's 3D coordinates with respect to RIS. We can obtain the UE's prediction bounding box and the UE's direction output by vision algorithm at a certain moment as shown in Fig.~\ref {positionofthereceiver.fig}. Then, according to UE's direction, RIS can select the optimal codeword among the pre-calculated codebook to refresh the reflection coefficients.  

In the real test, we move the UE back and forth at a fixed angular velocity $28^{\circ}/s$ within the coverage of the RIS beamforming. In addition, we set the initial codeword of the RIS as the codeword with the desired pitch angle $\theta_{rx}$ of $90^{\circ}$ and the desired azimuth angle $\varphi_{rx}$ of $0^{\circ}$. To highlight the effectiveness of vision, we conduct two experiments both with and without visual assistance. When there is no visual assistance, the traditional beam-sweeping method is used to achieve beam tracking \cite{9406940}. In the beam-sweeping process, the feedback procedure is ignored and the UE side is directly connected to the RIS. The specific beam-sweeping strategy is: firstly, the full-range beam-sweeping is used to find the optimal beam direction; then during the beam tracking process, if the signal-to-noise ratio (SNR) of the UE side is 6~dB lower than the maximum SNR, a small range of scanning is performed near the current beam direction.

\begin{figure}[!t]
	\centering
	\includegraphics[width=3.5in]{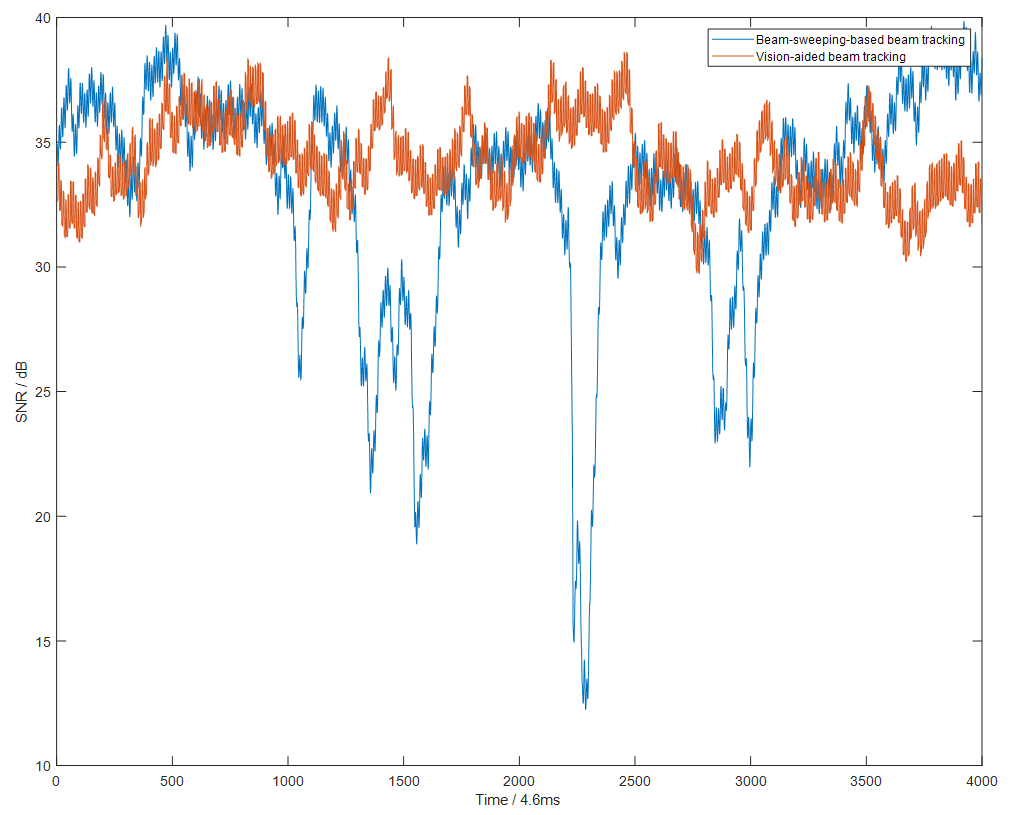}
	\caption{SNR variation curve of the UE under case \uppercase\expandafter{\romannumeral1}.}
	\label{snrscenario1_first_case}       
\end{figure}

We calculate the real-time SNR at the UE to evaluate the beamforming performance of the vision-aided RIS. The SNR variation curves are displayed in Fig.~\ref {snrscenario1_first_case}. It is seen that when the UE moves around, the SNR curve fluctuates steadily up and down around 35 dB with visual assistance, which ensures the high quality of the communication. Meanwhile, there is no additional feedback process and beam training overhead during the whole communication process. However, if the beam-sweeping method is adopted, there will be a significant drop in SNR at some moments, e.g., the interval of 9200 ms to 11500 ms in the Fig.~\ref {snrscenario1_first_case}. When the received SNR is poor, the RIS needs to perform beam sweeping to achieve beam tracking. However, normal communication is not possible during the beam-sweeping process, which introduces a certain beam training overhead.

\subsection{Test Under the Case \uppercase\expandafter{\romannumeral2}}
In case \uppercase\expandafter{\romannumeral2}, the RIS is used as an independent component of the communication system to assist the communication between the BS and the UE. Here we use a transmitting horn antenna at a distance of 3 meters from the RIS as the BS, and use a receiving horn antenna at a distance of 2.2 meters from the RIS as the UE. Meanwhile, the LOS path between BS and UE is blocked. The layout of the test scenario is shown in Fig.~\ref {snrscenario2_exp}. When the LOS path is blocked, the received SNR on the UE side is poor, which reduces the spectral efficiency significantly. The task for vision-aided RIS is to create another communication path between BS and UE to improve the SNR based on the visual information. 
Similar to the beam tracking scheme in case \uppercase\expandafter{\romannumeral1}, the vision-aided RIS needs to find the UE and calculate the exact UE's 3D coordinates with respect to the RIS, and then adjusts the reflection coefficients of the RIS according to the codebook in time. The difference is that the distance between the BS and RIS is farther, and thus RIS works in far-field condition. Hence the codebook needs to be redesigned according to \eqref{model_ris8}, \eqref{design_ris3}, and \eqref{design_ris1}, which is elaborated in Section \uppercase\expandafter{\romannumeral4}.

We assume that the UE moves back and forth at a fixed angular velocity $28^{\circ}/s$ in the region where the LOS path is blocked and the initial codeword of the RIS is set as the codeword with the desired pitch angle $\theta_{rx}$ of $90^{\circ}$ and the desired azimuth angle $\varphi_{rx}$ of $0^{\circ}$. Correspondingly, we conduct the communication test in both the situation of RIS with and without visual assistance. Similarly, in the absence of visual assistance, we use the beam-sweeping method to achieve beam tracking. We plot the obtained SNR variation curves in Fig.~\ref {snrscenario2_exp}. It can be seen that even if the LOS path between the BS and the UE is blocked, the vision-aided RIS can adjust another direct path to compensate for the performance degradation, keeping the SNR stable between 20 dB and 25 dB and without additional beam training overhead and feedback overhead. However, without the help of visual information, the SNR can fall below 15 dB at some moments, which raises the communication delay. In addition, when the beam-sweeping method is adopted, the receiving SNR jitter is more drastic compared to the case \uppercase\expandafter{\romannumeral1}. 
\begin{figure}[!t]
	\centering
	\includegraphics[width=3.5in]{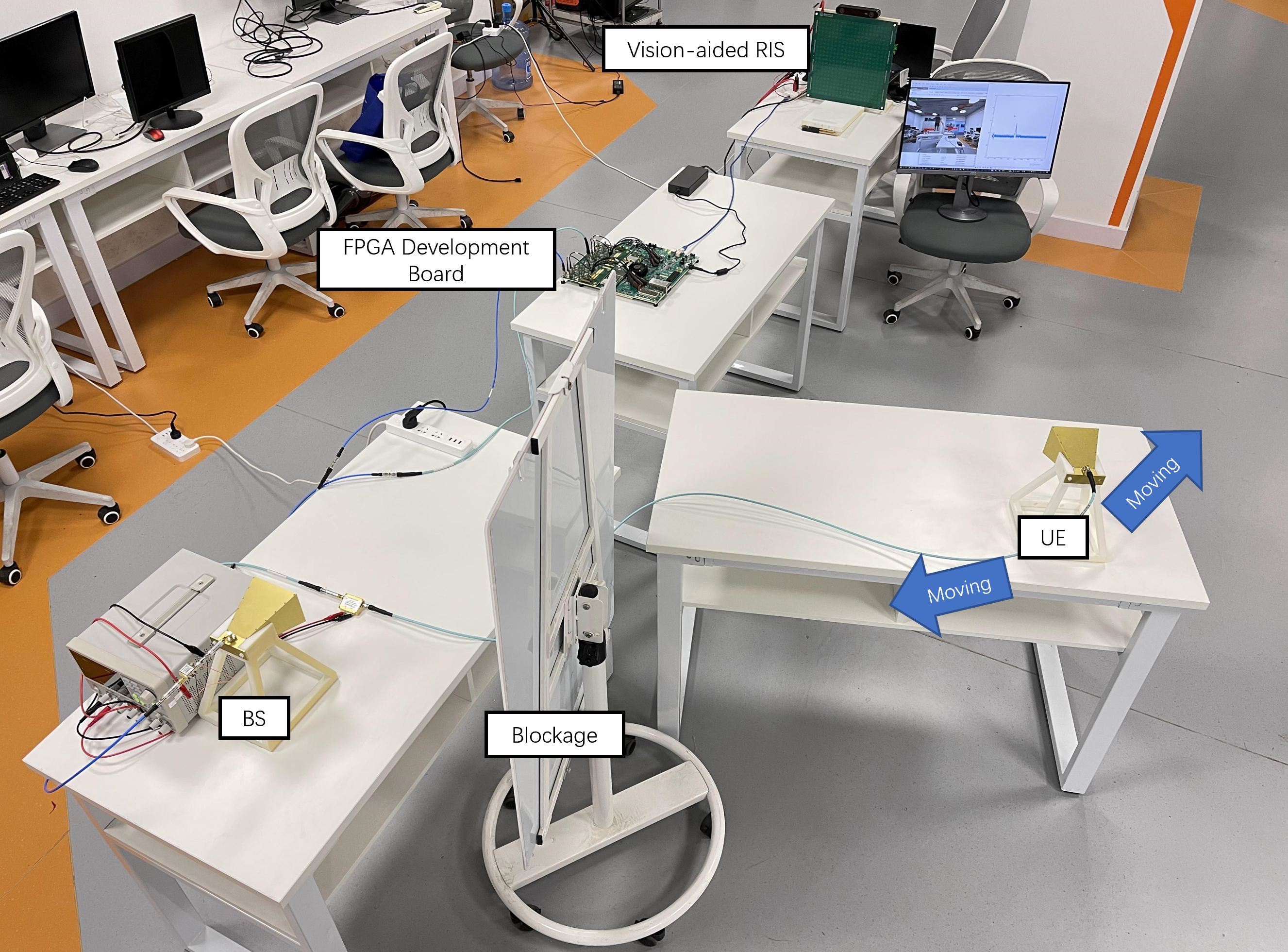}
	\caption{Test scenario layout under case \uppercase\expandafter{\romannumeral2}. In this scenario, we use a transmitting horn antenna at a distance of 2 meters from the RIS as the BS, and use a receiving horn antenna at a distance of 3 meters from the RIS as the UE. Meanwhile, the LOS path between BS and UE is blocked.}
	\label{SystemlayoutScenario2.fig}       
\end{figure}
\begin{figure}[!t]
	\centering
	\includegraphics[width=3.5in]{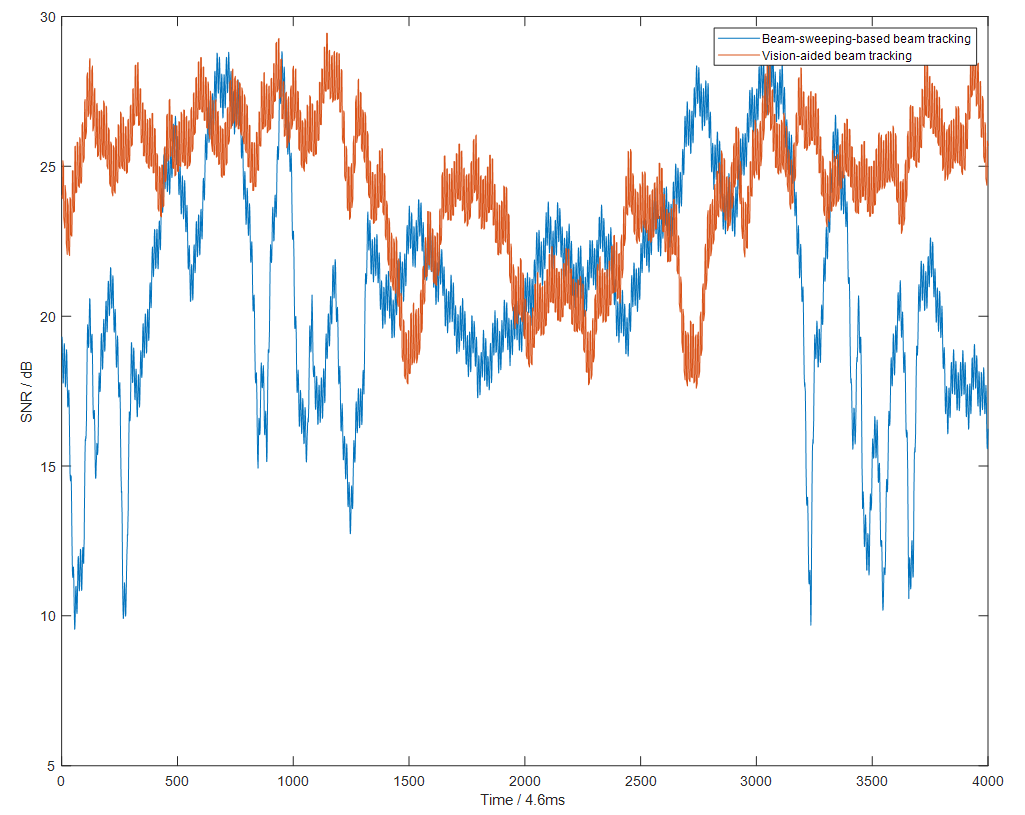}
	\caption{SNR variation curve of the UE under case \uppercase\expandafter{\romannumeral2}.}
	\label{snrscenario2_exp}       
\end{figure}

\section{Conclusion}
In this paper, we propose a novel computer vision-based approach to aid RIS for dynamic beam tracking and then implement the corresponding prototype verification system. A camera is attached at the RIS to obtain the visual information about the surrounding environment. With the object detection and binocular stereo vision algorithm, the vision-aided RIS can calculate the UE's 3D coordinates with respect to the RIS. With the UE's 3D coordinates, RIS quickly identifies the desired reflected beam direction and then adjusts the reflection coefficients according to the pre-designed codebook. Compared to the conventional approaches that utilize channel estimation or beam sweeping to design the reflection coefficients, the proposed approach not only saves beam training overhead but also eliminates the requirement for extra feedback links. Next we build a 20-by-20 RIS running at 5.4 GHz and develop a high-speed control board to ensure the real-time refresh of the reflection coefficients. Meanwhile we implement an independent peer-to-peer communication system to simulate the communication between the BS and the UE by referring to the 802.11n physical layer standard.

We calculate the radiation patterns by using the three-dimensional electromagnetic field simulation software CST and then compare the simulation results with the pratical test results in an anechoic chamber. Both simulation and test results of the radiation patterns show that the angle of the main lobe center of the radiation pattern is consistent with the desired angle, which verifies the correctness of the codebook design principle. Then we test the vision-aided RIS prototype system in two cases and compare it with the traditional beam-sweeping method. In case \uppercase\expandafter{\romannumeral1}, the RIS works as a passive array antenna at BS to achieve beamforming. In case \uppercase\expandafter{\romannumeral2}, the RIS is used as an independent component of the communication system to assist the communication between the BS and the UE and the LOS path between the BS and the UE is blocked. Both experimental results show that the vision-aided RIS can achieve real-time beam tracking and stabilize the received SNR of the UE around the normal values. Meanwhile, the proposed RIS beam tracking method does not require any beam training overhead and feedback overhead.

In the proposed prototype system, thanks to the high speed control board, the time to refresh the RIS codeword does not exceed 100 us, and thus the delay of beam tracking depends entirely on the calculation time of the UE's 3D coordinates (about 85 ms). Meanwhile, it is can be seen from Fig.~\ref {rispattern_fig} that the main lobe width of the RIS beam is greater than 10 degree. The proposed vision-aided RIS enables real-time beam tracking as long as the UE's angular velocity does not exceed 118 $^{\circ}/s$, which is much higher than the angular velocity of people or cars in real life. Therefore, the proposed vision-aided RIS in this paper could be widely applicable to various communication systems.

\bibliographystyle{IEEEtran}
\bibliography{ref}

\end{document}